\documentclass[aps,twocolumn,pra,superscriptaddress,showpacs,showkeys,floatfix]{revtex4-1}
\usepackage{graphicx,color}% Include figure files
\usepackage{epsfig}
\usepackage{graphicx}
\usepackage{amsmath,amssymb}
\usepackage{mathrsfs}
\usepackage{color}
\usepackage{bm}%

 \newcommand{\bra}[1]{\langle{#1} |}
 \newcommand{\ket}[1]{|{#1}\rangle  }
 
 \newcommand{\ketbra}[2]{\vert {#1} \rangle \langle{#2}\vert}

\providecommand{\openone}{\leavevmode\hbox{\small1\kern-3.8pt\normalsize1}}

\begin{document}

\title{Classical and quantum capacities of a fully 
correlated amplitude damping channel}

\author{A. D'Arrigo}
\affiliation{CNR-IMM UOS Universit\`a (MATIS), Consiglio Nazionale delle Ricerche, 
Via Santa Sofia 64, 95123 Catania, Italy}
\affiliation{Dipartimento di Fisica e Astronomia,
Universit\`a degli Studi Catania, Via Santa Sofia 64, 95123 Catania, Italy}
\author{G. Benenti} 
\affiliation{CNISM and Center for Nonlinear and Complex Systems,
Universit\`a degli Studi dell'Insubria, Via Valleggio 11, 22100 Como, Italy}
\affiliation{Istituto Nazionale di Fisica Nucleare, Sezione di Milano,
via Celoria 16, 20133 Milano, Italy}
\author{G. Falci}
\affiliation{CNR-IMM UOS Universit\`a (MATIS), Consiglio Nazionale delle Ricerche, 
Via Santa Sofia 64, 95123 Catania, Italy}
\affiliation{Dipartimento di Fisica e Astronomia,
Universit\`a degli Studi Catania, Via Santa Sofia 64, 95123 Catania, Italy}
\affiliation{Istituto Nazionale di Fisica Nucleare, Sezione di Catania, Via Santa
Sofia 64, 95123 Catania, Italy}
\affiliation{Centro Siciliano di Fisica Nucleare e Struttura della Materia, 
Via Santa Sofia 64, 95123 Catania, Italy}
\author{C. Macchiavello}
\affiliation{Dipartimento di Fisica and INFN-Sezione di Pavia, Via Bassi 
6, I-27100 Pavia, Italy}

\begin{abstract}  
We study information transmission over a fully correlated amplitude damping 
channel acting on two qubits. We derive the single-shot classical channel 
capacity and show that entanglement is needed to achieve the channel
best performance.
We discuss the degradability properties of the channel and evaluate the 
quantum capacity for any value of the noise parameter. We finally compute the
entanglement-assisted classical channel capacity. 
\end{abstract}                                                                 

%03.67.-a Quantum information,
%03.67.Hk Quantum communication, 
%03.65.Yz Decoherence; open systems; quantum statistical methods

\maketitle

\section{Introduction}

\label{sec:intro}

Physical processes can be viewed,
in terms of information theory,
as channels mapping the input (initial) state
onto the final (output) state, the transmission being 
in space (as in communication channels) or in time 
(as in the run of a computer).  
The performance of a noisy classical channel
can be characterized by a single number, i.e., its \textit{capacity},
defined as the maximum rate at which information can be reliably 
transmitted down the channel~\cite{cover-thomas}. 
On the other hand, noisy quantum communication 
channels~\cite{nielsen-chuang,benenti-casati-strini} can use
quantum systems as carriers of both classical or quantum information,  
by encoding classical bits by means of quantum states or by transferring 
(unknown) quantum states between, say, subunits of a quantum computer.
Therefore, different capacities must be defined.
The \textit{classical capacity} 
$C$~\cite{hausladen,schumacher-westmoreland,holevo98} and the 
\textit{quantum capacity} $Q$~\cite{lloyd,barnum,devetak} of a noisy quantum
channel are defined as the maximum number of, respectively, 
bits and qubits that can be reliably transmitted 
per channel use. The \textit{entanglement-assisted classical capacity}
$C_E$ gives the capacity of transmitting classical information,
provided the sender and the receiver share unlimited prior
entanglement~\cite{adami-Cerf,bennett1999,bennett-shor}. 
This quantity upper bounds the other capacities: we have
$Q\le C\le C_E$~\cite{footnote}.

Noise effects can be conveniently described in the quantum 
operation formalism~\cite{nielsen-chuang,benenti-casati-strini}:
any input state $\rho$ is mapped onto the output state 
$\rho'=\mathcal{E}(\rho)$ by a linear, completely positive, trace 
preserving (CPT) map $\mathcal{E}$. The simplest 
models for quantum channels are memoryless, that is, 
the quantum operation describing $n$ channel uses is
$\mathcal{E}_n=\mathcal{E}^{\otimes n}$. On the other hand, 
real systems exhibit \textit{memory} -or \textit{correlation}- effects among
consecutive uses. Such effects become unavoidable when 
increasing the transmission rate in quantum channels, 
as it can be explored experimentally in optical fibers~\cite{banaszek},
or in solid-state implementations of quantum hardware suffering
from low-frequency noise~\cite{solid-state}.
Quantum memory channels~\cite{memo_review}, i.e.
$\mathcal{E}_n\ne \mathcal{E}^{\otimes n}$,
attracted increasing attention in the last years.
Interesting new features emerge in several models,
including depolarizing channels~\cite{mp02,MMM}, 
Pauli channels~\cite{mpv04,daems,dc}, dephasing 
channels~\cite{hamada,njp,virmani,gabriela,lidar},
Gaussian channels~\cite{cerf},  
lossy bosonic channels~\cite{mancini,lupo}, 
spin chains~\cite{spins}, collision models~\cite{collision},
complex network dynamics~\cite{caruso},
and a micromaser model~\cite{micromaser}. 
In particular, phenomenological models with Markovian
correlated noise
(see, e.g., \cite{mp02,mpv04,bm04,daems,hamada,njp,virmani,datta,lmm09}) 
show that the transmission of classical information
can be enhanced by employing maximally entangled rather than 
separable states as information carriers~\cite{mp02,mpv04,daems}.
Furthermore, memory can enhance the quantum 
capacity of a channel, as shown for a Markov-chain dephasing 
channel, whose quantum capacity can be analytically 
computed~\cite{njp,virmani}. The main difficulty in the 
calculation of quantum channel capacities resides in the fact
that, due to the super-additivity property of the related 
entropic quantities~\cite{barnum,hastings}, maximization is requested over
all possible $n$-use input states, in the limit $n\to\infty$.  
For this reason, so far only a few memory channel models have
been fully solved in terms of their 
capacities~\cite{njp,virmani,lupo}. 

In this paper, we extend the class of solved quantum channels
to systems with damping, by considering a two-qubit amplitude
damping channel ${\cal E}_m$ with memory, in which the relaxation processes from a qubit
excited state towards the ground state only occur simultaneously
for the two qubits.  
The channel is 
parametrized by $\eta$ which is the conditional probability that the 
system, once 
it is found with the two qubits both in their excited state, does not decay. 
This channel is the fully correlated limit of the amplitude damping
channel with memory introduced in Ref.~\cite{yeo} and recently
investigated in Ref.~\cite{Jahangir}.
For channel ${\cal E}_m$ we compute the single-shot capacity $C_1$,
that is, the classical capacity optimized over single uses of 
the two-qubit channel, the quantum capacity $Q$ and the 
entanglement-assisted classical capacity $C_E$. In particular, 
we show that the ensemble optimizing the capacity $C_1$ must 
contain entangled two-qubit input states. 

The paper is organized as follows. In Sec. \ref{sec:model} we first
introduce the channel model and describe the channel
covariance properties. 
In Sec. \ref{sec:classical-capacity}, we discuss how to find
the quantum ensemble which maximizes the \textit{Holevo quantity}, showing
the explicit form of such optimal ensemble. 
We derive the form of the product state capacity $C_1$ 
of ${\cal E}_m$ and prove that entangled states are necessary to achieve
the capacity. We finally give an analytical expression for $C_1$. 
In Sec. \ref{sec:quantum-capacity} we show that the channel is 
degradable when $\eta$ is inside a given range; we find 
the system density operator which maximizes the 
\textit{coherent information}, and we determine the quantum capacity of 
${\cal E}_m$ for all possible values of $\eta$.
In Sec. \ref{sec:entass-capacity} we derive the entanglement-assisted 
channel capacity and we finally summarize the main results in Sect.
\ref{sec:conc}.

\section{The Model}
\label{sec:model}
We will first briefly review the memoryless amplitude damping 
channel (\textit{ad})~\cite{nielsen-chuang,benenti-casati-strini}, which 
acts on a generic single-qubit state $\rho$ as follows
\begin{equation}
 \rho \quad \rightarrow \quad \rho'={\cal E}_1(\rho)\,=\,\sum_{i \in \{0,1\}} E_i \,\rho\,E_i^\dag,
 \label{eq:ampl-damping-channel}
\end{equation}
where the Kraus operators $E_i$ are given by
\begin{equation}
{E}_0 = \left(
\begin{array}{cc}
  1 & 0   \\
  0 & \sqrt{\eta}  \\
  \end{array} \right), \qquad
{E}_1 = \left(
\begin{array}{cccc}
    0 & \sqrt{1-\eta}  \\
    0 & 0  \\
  \end{array} \right).
\label{eq:Memoryless-Kraus-Operators}
\end{equation}
Here we are using the orthonormal basis $\{\ket{0},\ket{1}\}$ 
($\sigma_z=\ketbra{0}{0}-\ketbra{1}{1}$). 
This channel describes relaxation processes, such as spontaneous 
emission of an atom, in which the system decays from the excited 
state $\ket{1}$ to the ground state $\ket{0}$. 
%The channel is parametrised by a single noise parameter $\eta\in[0,1]$.
The channel acts as follows on a generic single-qubit state
\begin{equation}
  \rho= 
    \left(
      \begin{array}{cc}
        1-p & \gamma   \\
        \gamma* & p  \\
      \end{array} 
    \right) \quad \rightarrow \quad
  \rho'= {\cal E}(\rho)=
    \left(
      \begin{array}{cc}
        1-\eta \,p & \sqrt{\eta}\,\gamma   \\
        \sqrt{\eta}\, \gamma* & \eta\, p  \\
      \end{array} 
    \right). 
\end{equation}
Note that the noise parameter $\eta$ ($0\le \eta\le 1$) plays the role of 
channel transmissivity. 
Indeed for $\eta=1$ we have a noiseless channel,
% since any state undergoes the channel without any distortion, 
whereas for $\eta=0$ the channel cannot carry any information since for any 
possible input we always obtain the same output state $\ket{0}$.

\begin{figure}[t!]
  \begin{center}
  \includegraphics[width=8.5cm]{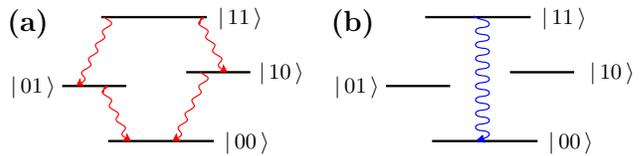}
  \end{center}
  \caption{Simple sketch of the relaxation mechanisms in the 
   channels ${\cal E}_1^{\otimes 2}$ (a) and ${\cal E}_m$ (b).
   In the memoryless setting ${\cal E}_1^{\otimes 2}$ relaxation
   is allowed from any level. In the full memory, relaxation phenomena
   in the two qubits are fully correlated, and relaxation is allowed
   only from $\ket{11}$.}
  \label{fig:scheme}
\end{figure}

For two channel uses, a \textit{memory} 
amplitude damping channel was introduced in 
Ref.~\cite{yeo}:
\begin{equation}
 \rho \quad \rightarrow \quad \rho'={\cal E}(\rho)\,=\,
(1-\mu) {\cal E}_1^{\otimes 2}(\rho)+\mu{\cal E}_m(\rho).
\label{eq:model}
\end{equation}
Here, $\rho$ is a generic two-qubit input state, and
$\mu$ ($0\le \mu\le 1$) is the memory parameter: the 
memoryless channel is recovered when $\mu=0$, while for 
$\mu=1$ we obtain the ``full memory'' amplitude 
damping channel ${\cal E}_m$.
In ${\cal E}_m$ the relaxation phenomena are fully correlated.
In other words,
when a qubit undergoes a relaxation process, the other qubit does the same, see Fig.~\ref{fig:scheme}.
In this way only the state $\ket{11}\equiv\ket{1}\otimes\ket{1}$  
can decay, while the other states $\ket{ij}\equiv\ket{i}\otimes\ket{j}$, $i,j\,\in\,\{0,1\}$,
$ij \neq 11$, are noiseless. 
In the Kraus formalism we have that
\begin{equation}
 \rho \quad \rightarrow \quad \rho'={\cal E}_m(\rho)\,=\,\sum_i B_i \,\rho\,B_i^\dag,
 \label{eq:full-memory-channel}
\end{equation}
with the Kraus operators 
%\small
\begin{equation}
{B}_0 = \left(
\begin{array}{cccc}
  1 & 0 & 0 & 0  \\
  0 & 1 & 0 & 0  \\
  0 & 0 & 1 & 0  \\
  0 & 0 & 0 & \sqrt{\eta}  \\
  \end{array} \right), \,\,
{B}_1 = \left(
\begin{array}{cccc}
  0 & 0 & 0 & \sqrt{1-\eta}  \\
  0 & 0 & 0 & 0  \\
  0 & 0 & 0 & 0  \\
  0 & 0 & 0 & 0  \\
  \end{array} \right).
\label{eq:memory-Kraus-Operators}
\end{equation}

In this paper we will focus on the fully correlated channel
${\cal E}_m$, for which we will compute analytically the 
single-shot classical capacity $C_1$, the quantum capacity $Q$,
and the entanglement-assisted classical capacity $C_E$.

%\normalsize

\subsection{Channel properties}
\label{subsec:channel-properties}

In this section, we discuss the covariance properties of channel ${\cal E}_m$,
from which the properties of general channel can be derived, 
with respect to some unitary transformations.
We first consider the following ones
\begin{equation} 
 {\cal R}_1=\sigma_z \otimes \openone, \quad {\cal R}_2=\openone \otimes \sigma_z, \quad
 {\cal R}_3=\sigma_z \otimes \sigma_z.
 \label{eq:R-operators}
\end{equation}
The Kraus operator $B_0$ commutes with each ${\cal R}_i$,
since $B_0$ and ${\cal R}_i$ have a diagonal form: $B_0 {\cal R}_i={\cal R}_i B_0 $.
%
%\\Regarding the relation among $B_1$ and ${\cal R}_i$,
The operator $B_1$ commutes with ${\cal R}_3$ and anticommutes
with ${\cal R}_1$ and ${\cal R}_2$:
\begin{eqnarray} 
&& {\cal R}_1 B_1 \,=\, -B_1  {\cal R}_1,\quad
 {\cal R}_2 B_1\,=\, -B_1  {\cal R}_2,\,\nonumber\\	
&&\hspace{2cm} {\cal R}_3 B_1\,=\, B_1  {\cal R}_3.
\end{eqnarray}
The channel 
is \textit{covariant} with respect to
the unitaries ${\cal R}_i$, namely
\begin{equation} 
 {\cal E}_m({\cal R}_i \, \rho \, {\cal R}_i)= {\cal R}_i\, 
{\cal E}_m(\rho)\,{\cal R}_i.
 \label{eq:Ri-cov}
\end{equation}
Actually, we can see that
\begin{eqnarray} 
   &&\hspace{-0.5cm}{\cal E}_m({\cal R}_1 \, \rho \, {\cal R}_1)\,= \nonumber\\
                       &&= B_0 {\cal R}_1 \rho {\cal R}_1 B^\dag_0 \,+\,
                           B_1 {\cal R}_1 \rho {\cal R}_1 B^\dag_1 \nonumber\\
                       &&=   {\cal R}_1 B_0 \rho B^\dag_0 {\cal R}_1  \,+\,
                           \big(-{\cal R}_1 B_1\big) \rho \big(-B^\dag_1 {\cal R}_1\big) \nonumber\\
                       &&= {\cal R}_1 \big( B_0 \rho B^\dag_0\,+\,
                            B_1 \rho B^\dag_1\big) {\cal R}_1 = {\cal R}_1 \,{\cal E}_m(\rho)\,{\cal R}_1, 
   \label{eq:R1-cov}
\end{eqnarray}
where we used $B^\dag_0=B_0$ and 
${\cal R}_1 B^\dag_1=(B_1 {\cal R}_1)^\dag=(-{\cal R}_1B_1)^\dag=-B_1^\dag{\cal R}_1$. Covariance under ${\cal R}_2$ can be proved in the same way.
Since ${\cal R}_3$ commutes with both $B_0$ and $B_1$, covariance 
of the channel under ${\cal R}_3$ follows trivially.

Finally, we consider the SWAP operation
\begin{equation}
{\cal S}_\textrm{w}\equiv\ketbra{00}{00}\,+\,\ketbra{01}{10}\,+\,\ketbra{10}{01}\,+\,\ketbra{11}{11}.
\end{equation}
The action of this gate is to transform the state $\ket{01}$ into 
$\ket{10}$, and vice versa, while it leaves unchanged the states 
$\ket{00}$ and $\ket{11}$.
From the structure of the operators (\ref{eq:memory-Kraus-Operators}),
it is immediate to verify that ${\cal S}_\textrm{w}$ commutes 
with $B_0$ and $B_1$.
Therefore the channel ${\cal E}_m$ is covariant with respect to 
${\cal S}_\textrm{w}$, namely
\begin{equation}
{\cal E}_m({\cal S}_\textrm{w} \, \rho \, {\cal S}_\textrm{w})= 
{\cal S}_\textrm{w}\, {\cal E}_m(\rho)\,{\cal S}_\textrm{w}.
\label{eq:Swap-cov}
\end{equation}

\section{Classical capacity}
\label{sec:classical-capacity}
The classical capacity $C$ of a quantum channel concerns the ability of the 
channel to convey classical information. It measures the maximum amount of 
classical information that can be reliably transmitted down the channel 
per channel use. In computing the classical capacity, the full 
optimization over all entangled uses is generally required. 
In this section, we address the problem
of finding the capacity $C_1$~\cite{nielsen-chuang} 
of the fully correlated channel ${\cal E}_m$. 
To do this we have to to maximize the so called Holevo quantity 
$\chi$~\cite{nielsen-chuang,benenti-casati-strini,holevo73}
with respect to one use of the channel ${\cal E}_m$.
Given a quantum source $\{p_\alpha, \rho_\alpha\}$, which is described by the density 
operator $\rho=\sum_\alpha p_\alpha \rho_\alpha$, we are dealing with the following 
optimization problem~\cite{schumacher-westmoreland,holevo98,hausladen}:
\begin{eqnarray}
  &&C_1({\cal E}_m)=\max_{\{p_\alpha,\rho_\alpha\}}{\chi\big({\cal E}_m,\{p_\alpha,\rho_\alpha\}\big)}, 
    \label{eq:C1} 
\end{eqnarray}
where the quantity to be optimized is the Holevo quantity, which is defined as
\begin{eqnarray}
  &&\hspace{-1cm}\chi\big({\cal E}_m,\{p_\alpha,\rho_\alpha\}\big) 
      \equiv S({\cal E}_m(\rho)) \, - \, \sum_\alpha p_\alpha S({\cal E}_m(\rho_\alpha)),
    \label{eq:chi}
\end{eqnarray}
where $S(\rho)=-{\rm Tr}(\rho\log_2 \rho)$ 
is the von Neumann entropy.
The first term in (\ref{eq:chi}) is the channel output entropy of
the quantum source described by $\rho$, whereas the second term is 
the channel average
output entropy.
Since for any ensemble of mixed states one can find an ensemble of pure states
 described by same density operator, and
whose Holevo quantity (\ref{eq:chi}) is at least as 
large~\cite{schumacher-westmoreland}, %(pag. 136, eq.(46-51)),
in the following we will only consider ensembles of pure states 
$\{p_k,\ket{\psi_k}\}$:
\begin{eqnarray}
  &&\hspace{-0.5cm} C_1({\cal E}_m)=\max_{\{p_k,\ket{\psi_k}\}}{\chi\big({\cal E}_m,\{p_k,\ket{\psi_k}\}\big)},
    \label{eq:C1-b} \\
  &&\hspace{-0.5cm}\chi\big({\cal E}_m,\{p_k,\ket{\psi_k}\}\big)\,= \nonumber \\
  &&\hspace{0.5cm}=\,S({\cal E}_m(\rho)) \, - \, \sum_k p_k S({\cal E}_m(\ketbra{\psi_k}{\psi_k})),
    \label{eq:chi-pure}
\end{eqnarray}
where now $\rho=\sum_k\, p_k\ketbra{\psi_k}{\psi_k}$.

\subsection{Searching for ensembles that maximize $\chi$}

In this section, we will use the channel covariance properties discussed in 
Sec.~\ref{subsec:channel-properties} to find the form
of the ensembles $\{p_k,\ket{\psi_k}\}$ solving the maximization 
problem (\ref{eq:C1-b})-(\ref{eq:chi-pure}). 
%among all possible ensembles in the Hilbert space spanned by $\{\ket{ij}\}$, 
%$i,j \,\in\,\{0,1\}$. We will use arguments similar to those 
%used in~\cite{giovannetti,dorlas-morgan08}.
We proceed along three steps: steps I and II exploit the
covariance properties of channel ${\cal E}_m$, while step III
uses the specific structure of the eigenvalues of the output states. 
Finally in step IV, we give the form of the optimal ensemble, and
the expression of corresponding Holevo quantity.

\subsubsection{Step I: Exploiting channel covariance with respect to 
the operations ${\cal R}_i$}
\label{sec:StepI}

Given a generic ensemble $\{p_k,\ket{\psi_k}\}$, we build a new ensemble  
by replacing 
each state $\ket{\psi_k}$ in $\{p_k,\ket{\psi_k}\}$ by the set
\begin{eqnarray}
\{\ket{\psi_k},\, {\cal R}_1\ket{\psi_k},\,{\cal R}_2\ket{\psi_k},\,{\cal R}_3\ket{\psi_k}\},
\nonumber
\end{eqnarray}
each state occurring with probability $\tilde{p}_k=p_k/4$~\cite{dorlas-morgan08}. We refer to
this new ensemble as $\{\tilde{p}_k,\ket{\tilde{\psi}_k}\}$, and call 
$\tilde{\rho}=\sum_k \tilde{p}_k \ketbra{\tilde{\psi}_k}{\tilde{\psi}_k}$ 
the associated density operator:
%We observe that
\begin{eqnarray}
   %  &&\hspace{-1.0cm} \rho = \sum_k p_k \ketbra{\psi_k}{\psi_k} \rightarrow \nonumber\\
     &&\hspace{-0.5cm} \tilde{\rho}\,=\,\sum_k \frac{p_k}{4} \Big(\ketbra{\psi_k}{\psi_k}\,+\,
                             \sum_{i=1}^3{\cal R}_i\ketbra{\psi_k}{\psi_k}{\cal R}_i\Big)\,=\nonumber\\
     &&\hspace{-0.1cm}       =\frac{1}{4}\Big(\rho \,+\,\sum_{i=1}^3{\cal R}_i\rho{\cal R}_i\Big).
    \label{eq:rho-tilde-a}
\end{eqnarray}
%and 
It can be verified that $\tilde{\rho}$ %(\ref{eq:rho-tilde-a}) 
has the same diagonal elements of $\rho$,
with all vanishing off-diagonal entries.

We now show that
\begin{equation}
   \chi\big({\cal E}_m,\{\tilde{p}_k,\ket{\tilde{\psi}_k}\}\big) \ge \chi\big({\cal E}_m,\{p_k,\ket{\psi_k}\}\big).
\label{eq_obj1}
\end{equation}
To this end we first notice that
% which describes the new ensemble.
%By using to the identities (\ref{eq:Ri-cov}), it turns out that
\begin{eqnarray}
&&\hspace{-0.5cm}
S\big({\cal E}_m(\ketbra{\tilde{\psi}_k}{\tilde{\psi}_k})\big)\,=\,
S\big({\cal E}_m({\cal R}_i \ketbra{\psi_k}{\psi_k} {\cal R}_i)\big) \,=\,
 \nonumber\\
&&\hspace{0.5cm}S\big({\cal R}_i\,{\cal E}_m(\ketbra{\psi_k}{\psi_k})\,{\cal R}_i\big)\,=\,
S\big({\cal E}_m(\ketbra{\psi_k}{\psi_k})\big),
\label{equalities-a}
\end{eqnarray}
where we used eqs. (\ref{eq:Ri-cov}) and the fact that 
a unitary operation does not change the von Neumann entropy. 
Therefore, by replacing the old ensemble with the new one, 
the second term in the Holevo quantity 
(\ref{eq:chi-pure}) does not change: 
\begin{eqnarray}
      &&\sum_k \tilde{p}_k S({\cal E}_m(\ketbra{\tilde{\psi}_k}{\tilde{\psi}_k}))
       =4 \sum_k \frac{p_k}{4} S({\cal E}_m(\ketbra{\psi_k}{\psi_k}))\nonumber\\ 
      &&\hspace{2cm}= \sum_k p_k S({\cal E}_m(\ketbra{\psi_k}{\psi_k})).
    \label{eq:chi-second-term-a}
\end{eqnarray}
%Let us consider the changes in the first term of the Holevo quantity.

For the output entropy related to $\tilde{\rho}$ we find:
\begin{eqnarray}
     &&\hspace{-0.5cm} S({\cal E}_m(\tilde{\rho}))\,=\,
            S\Big({\cal E}_m\Big(\frac{1}{4}\rho \,+\,\frac{1}{4}\sum_{i=1}^3{\cal R}_i\rho{\cal R}_i \Big)\Big)\nonumber\\
     &&\hspace{-0.0cm}\,=\,
            S\Big(\frac{1}{4}{\cal E}_m\big(\rho\big) \,+\,
                 \frac{1}{4}\sum_{i=1}^3{\cal E}_m\big({\cal R}_i\rho{\cal R}_i\big)\Big) \,  \nonumber\\
     &&\hspace{-0.0cm} \ge 
            \frac{1}{4}S\big({\cal E}_m\big(\rho\big)\big) \,+\,
               \frac{1}{4}\sum_{i=1}^3 S\big({\cal E}_m\big({\cal R}_i\rho{\cal R}_i\big)\big) \nonumber\\
     &&\hspace{+0.0cm} = S\big({\cal E}_m\big(\rho\big)\big),
     \label{eq:chi-first-term-a}
\end{eqnarray}
where we have used the linearity of %the quantum operation
 ${\cal E}_m$, the concavity of von Neumann 
entropy~\cite{nielsen-chuang}, and  Eq.~(\ref{equalities-a}).
%Eqs.~ (\ref{eq:chi-second-term-a}) 
%and (\ref{eq:chi-first-term-a}) imply that
%\begin{equation}
%   \chi\big({\cal E}_m,\{\tilde{p}_k,\ket{\tilde{\psi}_k}\}\big) \ge \chi\big({\cal E}_m,\{p_k,\ket{\psi_k}\}\big).
%\end{equation}

Relations (\ref{eq:chi-second-term-a}) 
and (\ref{eq:chi-first-term-a}) prove the inequality~(\ref{eq_obj1}), and we can summarize 
the conclusions as follows:
%As a conclusion of the above discussion we can state that 
for any ensemble 
of pure states we can find another ensemble,
whose density matrix has the same diagonal, with zero off-diagonal 
entries, and whose Holevo quantity is at least as large.
In the following, we will work with ensembles
$\{\tilde{p}_k,|{\tilde\psi}_k\rangle\}$,
we will omit the tilde hereafter.

To fix the notation, we introduce the expression of the generic input state in  
$\{{p}_k,|{\psi}_k\rangle\}$:
\begin{equation}
    \ket{\psi_k}\,=\,a_k \ket{00}\,+\,b_k\ket{01}\,+\,c_k\ket{10}\,+\,d_k\ket{11},
  \label{eq:generic-input-state}
\end{equation}
where  $a_k,\,b_k,\,c_k,\,d_k\ \in \mathbb{C}$ and 
 $|a_k|^2+|b_k|^2+|c_k|^2+|d_k|^2=1$.
The corresponding density matrix is given by
\begin{equation}
{\rho}\,=\,\sum_k p_k\ketbra{\psi_k}{\psi_k} = \left(
\begin{array}{cccc}
  \alpha & 0 & 0 & 0  \\
  0 & \beta & 0 & 0  \\
  0 & 0 & \gamma & 0  \\
  0 & 0 & 0 & \delta  \\
  \end{array} \right),
\label{eq:diagonal-density-operator} 
\end{equation}
where
\begin{eqnarray}
&&\alpha=\sum_k p_k |a_k|^2, \quad \beta=\sum_k p_k |b_k|^2,
\quad \gamma=\sum_k p_k |c_k|^2, \nonumber\\
&&\hspace{1cm} \delta=\sum_k p_k |d_k|^2\,=\,1\,-\,\alpha\,-\,\beta\,-\,\gamma.
\label{eq:populations} 
\end{eqnarray}

\subsubsection{Step II: Exploiting channel covariance with respect to 
the SWAP operation}
\label{sec:StepII}

%At this stage, we are dealing with the ensembles $\{p_k,\ket{\psi_k}\}$ 
%defined in 
%Eqs.~(\ref{eq:generic-input-state})-(\ref{eq:diagonal-density-operator}). 
%with real coefficients for $\ket{\psi_k}$. 

Starting from any ensemble $\{p_k,\ket{\psi_k}\}$ defined in 
Eqs.~(\ref{eq:generic-input-state})-(\ref{eq:diagonal-density-operator}), we 
can generate another ensemble,
by replacing each state $\ket{\psi_k}$ by the couple of states
$\{\ket{\psi_k},\,{\cal S}_\textrm{w}\ket{\psi_k}\}$, each one occurring with
probability $p_k/2$. 
The state ${\cal S}_\textrm{w}\ket{\psi_k}$ is obtained
from $\ket{\psi_k}$ by exchanging the coefficients $b_k$ and $c_k$ 
in eq.~(\ref{eq:generic-input-state}).
We call this new ensemble $\{\tilde{p}_k,\ket{\tilde{\psi}_k}\}$,
and $\tilde{\rho}$ the corresponding
density operator: % given by
\begin{eqnarray}
 %    &&\hspace{-0.5cm} \rho = \sum_k p_k \ketbra{\psi_k}{\psi_k} \rightarrow \nonumber\\
     &&\hspace{+0.0cm}\tilde{\rho}  \,=\,\frac{1}{2}\sum_k p_k\, \ketbra{\psi_k}{\psi_k}\,+
            \,\frac{1}{2}\sum_k p_k\, 
                         {\cal S}_\textrm{w}\ket{\psi_k}\bra{\psi_k}{\cal S}_\textrm{w} \nonumber\\
     &&\hspace{+0.3cm} =\,
      \frac{1}{2}\rho\,+\,\frac{1}{2}{\cal S}_\textrm{w}\rho{\cal S}_\textrm{w}
       \,=\, \frac{1}{2}\rho\,+\,\frac{1}{2}\rho(\beta\leftrightarrow \gamma). 
    \label{eq:rho-tilde-b}
\end{eqnarray}

Again the ensemble $\{\tilde{p}_k,\ket{\tilde{\psi}_k}\}$
has a Holevo quantity $\chi$ at least as large as that of the parent ensemble 
$\{{p}_k,\ket{{\psi}_k}\}$. 
\\To prove this we first observe
that the second term of $\chi$ eq.~(\ref{eq:chi-pure}) does not 
change:
\begin{eqnarray}
      &&\hspace{0cm}\sum_k \tilde{p}_k S({\cal E}_m(\ketbra{\tilde{\psi}_k}{\tilde{\psi}_k}))
       =\frac{1}{2}\sum_k p_k S({\cal E}_m(\ketbra{\psi_k}{\psi_k}))+\nonumber\\
       &&\hspace{1cm} +\,\frac{1}{2}\sum_k p_k S({\cal E}_m({\cal S}_\textrm{w}
        \ketbra{\psi_k}{\psi_k}{\cal S}_\textrm{w}))=\nonumber\\
      &&\hspace{0cm}=\sum_k p_k S({\cal E}_m(\ketbra{\psi_k}{\psi_k})),
    \label{eq:chi-second-term-b}
\end{eqnarray}
where we have used (\ref{eq:Swap-cov}). 
%One could also observe that 
%exchanging $b_k$ with $c_k$ does not produce any change in the eigenvalues
%(\ref{eq:eigenvalues-output-purestateE2}). 
%With regard to the first term of the Holevo quantity  (\ref{eq:chi-pure}), 
%we observe that:
Then for the first term we find:
%where ${\cal S}_\textrm{w}\rho{\cal S}_\textrm{w}$ is given by 
%(\ref{eq:diagonal-density-operator}) by exchanging $\beta$ with $\gamma$.
%Hence, the first term of the Holevo quantity (\ref{eq:chi-pure}) is 
\begin{eqnarray}
     &&\hspace{-0.4cm} S({\cal E}_m(\tilde{\rho}))\,=\,
            S\Big({\cal E}_m\Big(\frac{1}{2}\rho \,+\,\frac{1}{2}{\cal S}_\textrm{w}\rho{\cal S}_\textrm{w}\Big)\Big)\nonumber\\
&&\hspace{0.0cm}
            =S\Big(\frac{1}{2}{\cal E}_m\big(\rho\big) \,+\,\frac{1}{2}{\cal E}_m\big({\cal S}_\textrm{w}\rho{\cal S}_\textrm{w}\big)\Big) \nonumber\\
     &&\hspace{+0.0cm} \ge 
            \frac{1}{2}S\big({\cal E}_m\big(\rho\big)\big) \,+\,
            \frac{1}{2}S\big({\cal E}_m\big({\cal S}_\textrm{w}\rho{\cal S}_\textrm{w}\big)\big)\nonumber\\
&&\hspace{+0.0cm} = \frac{1}{2}S\big({\cal E}_m\big(\rho\big)\big) \,+\,
            \frac{1}{2}S\big({\cal S}_\textrm{w}{\cal E}_m\big(\rho\big){\cal S}_\textrm{w}\big)
\,=\,S\big({\cal E}_m\big(\rho\big)\big),
    \label{eq:chi-first-term-b}
\end{eqnarray}
where we have used arguments similar to those exploited in deriving
(\ref{eq:chi-first-term-a}), together with the covariance 
property (\ref{eq:Swap-cov}).
Relations (\ref{eq:chi-second-term-b}) and (\ref{eq:chi-first-term-b}) prove
the upper bound provided by $\chi\big({\cal E}_m,\{\tilde{p}_k,\ket{\tilde{\psi_k}}\}\big)$. 
%the new ensemble has a Holevo quantity at least as great as the old one.

\subsubsection{Step III: Exploiting the structure of the output state
eigenvalues}
\label{sec:StepIII}

When the channel ${\cal E}_m$ acts on the generic
state (\ref{eq:generic-input-state}), it yields an output given by
\small
\begin{equation}
{\rho}_k' = \left(
\begin{array}{cccc}
  |a_k|^2 +(1-\eta)|d_k|^2 & a_k b_k^*           & a_k c_k^*           & \sqrt{\eta}a_k d_k^*  \\
  a_k^* b_k                & |b_k|^2             & b_k c_k^*           & \sqrt{\eta}b_k d_k^*  \\
  a_k^* c_k                & b_k^*c_k            & |c_k|^2             & \sqrt{\eta}c_k d_k^*  \\
  \sqrt{\eta}a_k^* d_k     & \sqrt{\eta}b_k^*d_k & \sqrt{\eta}c_k^*d_k & \eta|d_k|^2  \\
  \end{array} \right).
\label{eq:output-purestateE2} 
\end{equation}
\normalsize
The above density operator has at least two zero eigenvalues, due to the fact
that the channel ${\cal E}_m$ has a noiseless subspace $\textrm{span}\{\ket{01},\ket{10}\}$ 
which does not mix with the other subspace $\textrm{span}\{\ket{00},\ket{11}\}$.
The remaining two eigenvalues are given by
\begin{eqnarray}
  \label{eq:eigenvalues-output-purestateE2}
   &&    l_{k\pm}\,=\,\frac{1}{2}\Big(1\,\pm\,\sqrt{1-z_k^2}\Big),\\
   &&     z_k^2=1-\big\{|a_k|^4 +2 |a_k|^2  (|b_k|^2 + |c_k|^2 + |d_k|^2) +\nonumber\\
   &&  \hspace{2cm}          \big[|b_k|^2 +  |c_k|^2 - |d_k|^2 (1 - 2 \eta)\big]^2\big\}. \nonumber
\end{eqnarray}
Since $l_{k\pm}$ do not depend on the phase of 
$a_k,\,b_k,\,c_k,\,d_k$, we can assume without loss of generality that these
 coefficients are real.
From (\ref{eq:eigenvalues-output-purestateE2}) the average
 output entropy is found
\begin{eqnarray}
   \sum_k p_k S({\cal E}_m(\ketbra{\psi_k}{\psi_k}))\,=\,
   \sum_k p_k H_2(l_k),
   \label{eq:average-output-entropy}
\end{eqnarray}
where $H_2(x)=-x\log_2(x)-(1-x)\log_2(1-x)$ is the Shannon binary entropy.
\\Starting from the mixed state (\ref{eq:diagonal-density-operator}) the map 
produces
\begin{equation}
{\rho}' \,=\,{\cal E}_m({\rho})\,=\, \left(
\begin{array}{cccc}
  \alpha +(1-\eta)\delta & 0      & 0      & 0            \\
        0                & \beta  & 0      & 0            \\
        0                & 0      & \gamma & 0            \\
        0                & 0      & 0      & \eta \delta  \\
  \end{array} \right),
\label{eq:output-diagonalstateE2} 
\end{equation}
and therefore the output entropy is given by
\begin{eqnarray}
    &&S({\cal E}_m({\rho}))\,=\,-[\alpha +(1-\eta)\delta]\log_2[\alpha +(1-\eta)\delta]\nonumber\\
    &&\hspace{1.5cm}      -\beta\log_2(\beta)-\gamma\log_2(\gamma)-\eta\delta\log_2(\eta\delta). 
   \label{eq:output-entropy}
\end{eqnarray}

We now modify the ensemble $\{\tilde{p}_k,\ket{\tilde{\psi}_k}\}$
introduced in Sec.~\ref{sec:StepII},
by replacing the coefficients $b_k$ and $c_k$ in each state 
$\ket{\tilde{\psi}_k}$
by $\overline{b}_k$ and $\overline{c}_k$, where 
$\overline{b}_k=\pm\,\overline{c}_k=\sqrt{(b_k^2+c_k^2)/2}$. %Il segno meno è necessario
%per imporre uguale a zero gli elementi fuori diagonale
We call this new ensemble
$\{\overline{p}_k,\ket{\overline{\psi}_k}\}$ and 
the corresponding density operator $\bar{\rho}$,
which is the same of $\tilde{\rho}$. Indeed
\begin{eqnarray}
      &&\hspace{0cm}\overline{\rho}=\sum_k \overline{p}_k \ketbra{\overline{\psi}_k}{\overline{\psi}_k}\nonumber\\
%      &&\hspace{0.2cm} =\sum_k \tilde{p}_k 
%          \ketbra{\tilde{\psi}_k(b_k\rightarrow\overline{b}_k,c_k\rightarrow\overline{c}_k)}
%                 {\tilde{\psi}_k(b_k\rightarrow\overline{b}_k,c_k\rightarrow\overline{c}_k)}\nonumber\\
      &&\hspace{0.2cm}
         = \left(
                                    \begin{array}{cccc}
                                      \sum_k \tilde{p}_k a_k^2 & 0                & 0                & 0       \\
                                          0  & \sum_k \tilde{p}_k\overline{b}_k^2 & 0                & 0       \\
                                          0  & 0                & \sum_k \tilde{p}_k\overline{c}_k^2 & 0       \\
                                          0  & 0                & 0                & \sum_k \tilde{p}_kd_k^2  \\
                                    \end{array} \right)\nonumber\\
      &&\hspace{0.2cm}
         = \sum_k \frac{p_k}{2}\left(
                                    \begin{array}{cccc}
                                      2a_k^2 & 0           & 0           & 0       \\
                                          0  & b_k^2+c_k^2 & 0           & 0       \\
                                          0  & 0           & b_k^2+c_k^2 & 0       \\
                                          0  & 0           & 0           & 2d_k^2  \\
                                    \end{array} \right)\nonumber\\
      &&\hspace{0.2cm}= \left(
                                  \begin{array}{cccc}
                                      \alpha & 0                         & 0                         & 0       \\
                                          0  & \frac{1}{2}(\beta+\gamma) & 0                         & 0       \\
                                          0  & 0                         & \frac{1}{2}(\beta+\gamma) & 0       \\
                                          0  & 0                         & 0                         & \delta  \\
                                    \end{array} \right)\nonumber\\
      &&\hspace{0.2cm} =\frac{1}{2}\rho\,+\,\frac{1}{2}\rho(\beta\leftrightarrow \gamma) =\tilde{\rho},
    \label{eq:chi-first-term-c}
\end{eqnarray}
where we have used relations (\ref{eq:populations}). The third equality 
comes from the fact that for any state $\ket{\tilde{\psi}_k}$
there is another one with the same occurrence probability $\tilde{p}_k=p_k/2$, which has the same $a_k, d_k$, 
but with $b_k$ exchanged with $c_k$.
It follows that the first term of the Holevo quantity is unchanghed.
This is true also for the second term. For 
$\{\overline{p}_k,\ket{\overline{\psi}_k}\}$ it reads:
\begin{eqnarray}
      &&\hspace{0cm}\sum_k \overline{p}_k S({\cal E}_m(\ketbra{\overline{\psi}_k}{\overline{\psi}_k}))
        =\sum_k \tilde{p}_k H_2(l'_k)
%                      \sum_k \tilde{p}_k S({\cal E}_m(\ketbra{\tilde{\psi}_k}{\tilde{\psi}_k})).
    \label{eq:chi-second-term-c}
\end{eqnarray}
and the new eigenvalues $l'_k$ are identical to the  
$l_{k\pm}$ in eq. (\ref{eq:eigenvalues-output-purestateE2}), since for
real coefficients, they
both depend on on the combination $b_k^2+c_k^2$
which is unaffected by the transformation
$b_k\rightarrow\overline{b}_k,c_k\rightarrow\overline{c}_k$.

 Therefore the ensemble 
$\{\overline{p}_k,\ket{\overline{\psi}_k}\}$ produces the same Holevo quantity
of $\{\tilde{p}_k,\ket{\tilde{\psi}_k}\}$ of Sec.~\ref{sec:StepII},
(as equations (\ref{eq:chi-first-term-c}) and (\ref{eq:chi-second-term-c}) 
  show), but has the advantage of 
a simpler structure of the states in the ensemble.
%\vspace{0.5cm}

\subsubsection{Step IV: optimal ensemble and the corresponding Holevo quantity}
\label{sec:StepIV}

The chain of relations obtained up to here proves that the ensemble 
$\{\bar{p}_k,\ket{\bar{\psi}_k}\}$ allows to achieve an upper bound
for the Holevo quantity of a generic ensemble $\{p_\alpha,\rho_\alpha\}$.
Indeed $\{\bar{p}_k,\ket{\bar{\psi}_k}\}$ belongs to the original 
ensemble $\{p_\alpha,\rho_\alpha\}$, the maximization of the Holevo quantity
for the former ensemble also yields the maximum over the whole set of 
$\{p_\alpha,\rho_\alpha\}$. 

Summing up and simplifying the notation,
we then have to explore
ensembles $\{{p}_k,\ket{{\psi}_k}\}$ 
($k \, \in \, \{1,2,\ldots ,N\}$), 
%which 
%maximizes the Holevo quantity for channel ${\cal E}_m$ has states
where states have the form
\begin{equation}
   \ket{\psi_k}\,=\,a_k \ket{00}\,+\,b_k\ket{01}\,\pm\,b_k\ket{10}\,+\,d_k\ket{11},
   \label{eq:optimal-ensemble-inputstate}
\end{equation}
with real $a_k, b_k, d_k$ ($a_k^2+2b_k^2+d_k^2=1$).
The density matrix of such ensemble has the form
\begin{equation}
\rho = \left(
\begin{array}{cccc}
  \alpha & 0 & 0 & 0  \\
  0 & \beta & 0 & 0  \\
  0 & 0 & \beta & 0  \\
  0 & 0 & 0 & \delta  \\
  \end{array} \right),
\label{eq:optimal-ensemble-density-operator} 
\end{equation}
where
\begin{eqnarray}
&&\hspace{0cm} \alpha=\sum_k p_k a_k^2, \quad \beta=\sum_k p_k b_k^2, \qquad\nonumber\\
&&\hspace{0.0cm} \delta=\sum_k p_k d_k^2=1-\alpha-2\beta.
\label{eq:optimal-populations} 
\end{eqnarray}
The output entropy is given by
\begin{eqnarray}
    &&\hspace{0cm}S({\cal E}_m(\rho))\,=\,-[\alpha +(1-\eta)\delta]\log_2[\alpha +(1-\eta)\delta]\,+\nonumber\\
    &&\hspace{2cm}-2\beta\log_2(\beta)
         -\eta\delta\log_2(\eta\delta).
   \label{eq:optimal-output-entropy}
\end{eqnarray}
The average output entropy reads
\begin{eqnarray}
  \sum_k p_k H_2\Big(\frac{1}{2}\Big(1+\sqrt{1-z_k^2}\Big)\Big),
   \label{eq:optimal-average-output-entropy}
\end{eqnarray}
where
\begin{eqnarray}
 z_k^2\,=\,4d_k^2(1-\eta)(2b_k^2+\eta d_k^2).
   \label{eq:z-optimal}
\end{eqnarray}
Finally, the Holevo quantity (\ref{eq:chi-pure}) is given by
\begin{eqnarray}
    &&\hspace{-0.2cm}\chi\big({\cal E}_m,\{p_k,\ket{\psi_k}\}\big) 
      \,=\, \label{eq:optimal-chi}\\
    &&\hspace{0.0cm}-[\alpha +(1-\eta)\delta]\log_2[\alpha +(1-\eta)\delta]\,+\nonumber\\
    &&\hspace{0.0cm}  -2\beta\log_2(\beta)\,-\,\eta\delta\log_2(\eta\delta)\,+\nonumber\\
    &&\hspace{0.0cm}
         -\sum_k p_k H_2\bigg(\frac{1}{2}\Big[1+\sqrt{1-4d_k^2(1-\eta)(2b_k^2+\eta d_k^2)}\big]\bigg).\nonumber
\end{eqnarray}
In the following subsections we will compute the maximum of $\chi$ 
over two-qubit states, i.e., for single-use input states belonging to the 
class
(\ref{eq:optimal-ensemble-inputstate})-(\ref{eq:optimal-ensemble-density-operator}), 
thus deriving
the classical capacity $C_1$ for the channel ${\cal E}_m$.

\subsection{A lower bound for $C_1$}
\label{section:lower-bound}

In order to find a lower bound for $C_1$, it is sufficient 
to compute the Holevo quantity (\ref{eq:optimal-chi})
for an arbitrary ensemble. We choose ensembles 
%belonging to the class 
%(\ref{eq:optimal-ensemble-inputstate})-(\ref{eq:optimal-ensemble-density-operator})),
of the special form:
\begin{eqnarray}
    \label{eq:ensemble-lowerbound}
   \{{p}_k,\ket{{\psi}_k}\}\,=\, \{{p}_{\varphi k},\ket{{\varphi}_k}\} \,\cup \, 
                                  \{{p}_{\phi k},\ket{{\phi}_k}\}, 
\end{eqnarray}
where $\sum_k (p_{\varphi k}+p_{\phi k}) =1$, and such that 
 $\ket{\varphi_k}\in \textrm{span}\{\ket{01},\ket{10}\}$, whereas 
 $\ket{\phi_k}\in \textrm{span}\{\ket{00},\ket{11}\}$.

From (\ref{eq:output-purestateE2}) it is clear that the transmission of
the states $\ket{\varphi_k}$ is noiseless 
($S({\cal E}_m(\ketbra{\varphi_k}{\varphi_k}))=0$), so that:
\begin{eqnarray}
   \hspace{-0.5cm}\sum_k p_k S({\cal E}_m(\ketbra{\psi_k}{\psi_k}))\,=\,
   \sum_k p_{\phi k} S({\cal E}_m(\ketbra{\phi_k}{\phi_k})).
   \label{eq:lowerbound-outputentropy-1}
\end{eqnarray}
It is worth noting that, since the subspace spanned by 
$\ket{01}$ and $\ket{10}$ is noiseless for the channel 
${\cal E}_m$, we can choose for the subensemble
$\{{p}_{\varphi k},\ket{{\varphi}_k}\}$
any pair of mutually orthogonal states
\begin{eqnarray}
&&\hspace{-0.5cm}
   \left\{
   \begin{array}{ll}
    p_{\varphi+}=\beta, &  \ket{\varphi_+}=\cos\theta\ket{01}\,+\,\sin\theta\ket{10},\\ \\
    p_{\varphi-}=p_{\varphi+}, &  \ket{\varphi_-}=\sin\theta\ket{01}\,-\,\cos\theta\ket{10}.\\
    \end{array}
     \right.
%    \label{eq:best-ensemble}
\end{eqnarray}
With this notation, the subensemble of separable states 
$\{(\beta,\ket{01}),(\beta,\ket{10})\}$ and the 
subensemble of maximally 
entangled states
$\{\beta,\frac{1}{\sqrt{2}}\,(\ket{01}\pm\ket{10}\}$
are recovered when $\theta=0$ and $\theta=\pi/4$, respectively. 
All values of $\theta$ give the same contribution 
$-2\beta\log_2(\beta)$ to the Holevo quantity (\ref{eq:optimal-chi}).
Therefore, as far as we consider ensembles (\ref{eq:ensemble-lowerbound}),
there is no advantage in using entangled input
states of $\textrm{span}\{\ket{01},\,\ket{10}\}$.
% as one expect from (\ref{eq:optimal-ensemble-inputstate}).

With regard to the subensemble $\{{p}_{\phi k},\ket{{\phi}_k}\}$,
it is interesting to examine two instances. 
First we choose a set of product states 
$\{(\alpha,\ket{00}),(\delta,\ket{11})\}$, 
calling $\cal A$ the corresponding
ensemble $\{p_k,\ket{\psi_k}\}$. In this case we obtain
\begin{eqnarray}
  &&\hspace{-0.2cm}\sum_k p_{\phi k} S({\cal E}_m(\ketbra{\phi_k}{\phi_k}))\,
   =\,\alpha S({\cal E}_m(\ketbra{00}{00}))\,+\nonumber\\
  &&\hspace{0.3cm}  +\,  \delta\, S({\cal E}_m(\ketbra{11}{11}))\,=\,
  \delta\, H_2(\eta),
\end{eqnarray}
since from (\ref{eq:output-purestateE2}) it turns out that 
${\cal E}_m(\ketbra{00}{00})=\ketbra{00}{00}$ and 
${\cal E}_m(\ketbra{11}{11})=(1-\eta)\ketbra{00}{00}+\eta\ketbra{11}{11}$. 
The Holevo quantity (\ref{eq:optimal-chi}) relative to the ensemble 
$\cal A$ is
\begin{eqnarray}
  &&\hspace{-0.5cm}\chi\big({\cal E}_m,{\cal A}\big)=
   -[\alpha +(1-\eta)\delta]\log_2[\alpha +(1-\eta)\delta]\nonumber\\
  &&\hspace{1.5cm}  -2\beta\log_2(\beta)-\eta\delta\log_2(\eta\delta)-\delta H_2(\eta),
  \label{eq:chi-A}
\end{eqnarray}
so that a lower bound to $C_1$ is provided by
\begin{eqnarray}
  &&\hspace{-0.5cm}\chi^{(lb)}_1=\max_{\alpha,\beta,\delta}\chi\big({\cal E}_m,{\cal A}\big),
  \label{eq:lowerbound1}
\end{eqnarray}
with $\alpha, \beta, \delta$ real and $\alpha+2\beta+\delta=1$.

Secondly, for the subensemble $\{{p}_{\phi k},\ket{{\phi}_k}\}$ we choose
a set of entangled states
\begin{equation}
p_{\phi_\pm}=\frac{\alpha+\delta}{2}, \;\;
           \ket{\phi_\pm}=\sqrt{\frac{\alpha}{\alpha+\delta}}\ket{00}
\pm\sqrt{\frac{\delta}{\alpha+\delta}}\ket{11},
    \label{eq:best-ensemble}
\end{equation}
%\begin{equation}
%\{(\alpha+\delta)/2,\sqrt{\alpha/(\alpha+\delta)}\ket{00}\pm\sqrt{\delta/(\alpha+\delta)} \ket{11}\},
%\label{eq:ensemble-B}
%\end{equation}
calling $\cal B$ the corresponding ensemble,
for which we have
\begin{eqnarray}
  &&\hspace{-0.5cm}\sum_k p_{\phi k} S({\cal E}_m(\ketbra{\phi_k}{\phi_k}))\,=\,\nonumber\\
  && (\alpha+\delta) H_2\Bigg(\frac{1}{2}\Bigg[1+\sqrt{1-4\eta(1-\eta) \Big(\frac{\delta}{\alpha+\delta}\Big)^2}\Bigg]\Bigg),
\end{eqnarray}
as the output states generated by ${\cal E}_m$ from the input states (\ref{eq:best-ensemble})
have the same entropy, see eq. (\ref{eq:eigenvalues-output-purestateE2}).
The Holevo quantity (\ref{eq:optimal-chi}) relative to the ensemble $\cal B$ is
\begin{eqnarray}
  &&\hspace{-0.3cm}\chi\big({\cal E}_m,{\cal B}\big)=
   -[\alpha +(1-\eta)\delta]\log_2[\alpha +(1-\eta)\delta]+\nonumber\\
  &&\hspace{-0.2cm}\,-2\beta\log_2(\beta)\,-\,\eta\delta\log_2(\eta\delta)+\nonumber\\
  &&\hspace{-0.2cm} -\,(\alpha+\delta) H_2\Bigg(\frac{1}{2}\Bigg[1+\sqrt{1-4\eta(1-\eta) \Big(\frac{\delta}{\alpha+\delta}\Big)^2}\Bigg]\Bigg),
  \label{eq:chi-B} 
\end{eqnarray}
yielding the lower bound for $C_1$ given by 
\begin{eqnarray}
  &&\hspace{-0.5cm}\chi^{(lb)}_2=\max_{\alpha,\beta,\delta}\chi\big({\cal E}_m,{\cal B}\big),
  \label{eq:lowerbound2}
\end{eqnarray}
%where, as before, $\alpha, \beta, \delta$ are real and fulfill the 
%constraint $\alpha+2\beta+\delta=1$.

\begin{figure}[t!]
  \begin{center}
  \includegraphics[width=7.5cm]{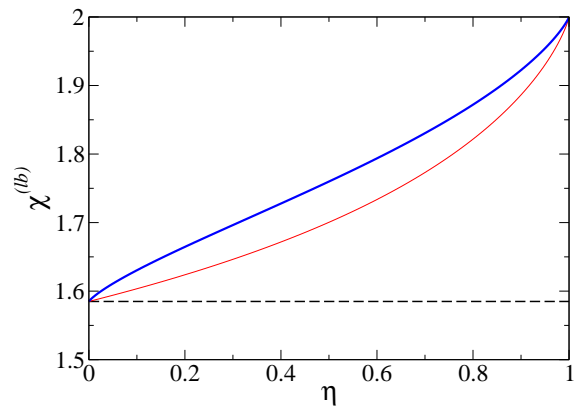}
  \end{center}
  \caption{(Color online) Maximum (obtained via numerical optimization)
 Holevo quantity relative to the ensembles $\cal A$
 (thin red curve) and $\cal B$ (thick blue curve) as a function of the 
channel transmissivity $\eta$. In the first case, we 
 obtain the lower bound $\chi^{(lb)}_1$ 
(\ref{eq:lowerbound1}) to the capacity $C_1$, 
 in the second the lower bound $\chi^{(lb)}_2$ (\ref{eq:lowerbound2}).
           We also plot the trivial lower bound
           $\log_2 3$ (dashed line).}
  \label{fig:bounds}
\end{figure}

We plot the bounds (\ref{eq:lowerbound1}) and (\ref{eq:lowerbound2}) in 
Fig.~\ref{fig:bounds}. Ensemble $\cal B$ 
(thick curve in the picture) always produces a 
better performance than ensemble $\cal A$ (thin curve). 
This result is a first hint that entangled input states 
may be useful to improve the 
channel capability to convey classical information.
Moreover, the classical 
capacity of ${\cal E}_m$ is at least equal to $\log_2(3)$,
reflecting the fact that in the worst case
($\eta=0$) there are three states allowing for noiseless transmission: 
$\ket{00},\ket{01},\ket{10}$. The lower bound $\log_2(3)$ is found by using 
them to encode three classical
symbols, each one occurring with the same probability $1/3$.
%It is worth noticing
%that in this case, the choice of the encoding is not only one: also encoding
%the three classical symbols by means of $\{\ket{11},\ket{01},\ket{10}\}$,
%produces the same bound, indeed when $\eta=0$ the transformation
%$\ket{11}\,\rightarrow\,\ket{00}$ is deterministic, so once the receiver
%obtain $\ket{00}$ it unambiguously infers the input state $\ket{11}$.

\subsection{The $C_1$ capacity of ${\cal E}_m$}
\label{section:c1-capacity}

Now we are ready to find the optimal ensemble, whose maximum Holevo quantity 
gives the $C_1$ classical capacity of ${\cal E}_m$.
To this end, 
we consider a generic ensemble  $\{{p}_k,\ket{{\psi}_k}\}$ belonging to 
the class 
(\ref{eq:optimal-ensemble-inputstate})-(\ref{eq:optimal-ensemble-density-operator}),
and we replace each state $\ket{\psi_k}$ and its occurrence probability $p_k$ 
in this ensemble by
\begin{eqnarray}
&&\hspace{-0.0cm} p_k,\quad\ket{\psi_k}\,\rightarrow\, \nonumber\\ \nonumber\\
&&\hspace{0.1cm}
   \left\{
   \begin{array}{ll}
    p_{\phi k}=\frac{p_k (a_k^2+d_k^2)}{2}, &
               \ket{\phi_{k\pm}}=\frac{a_k}{\sqrt{a_k^2+d_k^2}}\ket{00}\pm
               \frac{d_k}{\sqrt{a_k^2+d_k^2}}\ket{11},\nonumber\\ %\label{eq:states-1}
               \quad & \quad \\
               p_{\varphi k}=p_k b_k^2,&
               \ket{\varphi_{k\pm}}=\frac{1}{\sqrt{2}}\big(\ket{01}\pm e^{\pi i k/N}\ket{10}\big),
               \end{array} \right.\\
               \label{eq:states-2}
\end{eqnarray}
where the index $k$ ranges in $\{1,N\}$. We call 
$\{\tilde{p}_k,\ket{\tilde{\psi}_k}\}$ the new ensemble.
It is straightforward to prove that the density matrix of the new
ensemble is equal to that of the old ensemble 
(\ref{eq:optimal-ensemble-inputstate})-(\ref{eq:optimal-ensemble-density-operator}), 
and therefore the output entropy is unchanged: 
$S\big({\cal E}_m(\sum_k \tilde{p}_k \ketbra{\tilde{\psi}_k}{\tilde{\psi}_k})\big)=
 S\big({\cal E}_m(\sum_k p_k \ketbra{\psi_k}{\psi_k})\big)$. 

With regards to the second term
of the Holevo quantity, %(\ref{eq:chi-pure}) 
we notice that the states $\ket{\varphi_{k\pm}}$
in (\ref{eq:states-2}) do not contribute to the average output entropy:
$S({\cal E}_m(\ketbra{\varphi_{k\pm}}{\varphi_{k\pm}}))=0$.
%indeed the output entropy of any pure state belonging to subspace spanned
%by $\ket{01},\ket{10}$ vanishes, since its transmission through
%${\cal E}_m$ is noiseless. %(see (\ref{eq:output-purestateE2})). 
Therefore, the average entropy for the new ensemble is
\begin{eqnarray}
   &&\hspace{-0.0cm}\sum_k \tilde{p}_k S({\cal E}_m(\ketbra{\tilde{\psi}_k}{\tilde{\psi}_k}))\,=\,\nonumber\\
   &&\hspace{0cm}=\sum_k p_{\phi k} \Big(S({\cal E}_m(\ketbra{\phi_{k+}}{\phi_{k+}}))
                          +S({\cal E}_m(\ketbra{\phi_{k-}}{\phi_{k-}}))\Big)\nonumber\\
   &&\hspace{0cm}=2\sum_k  p_{\phi k} S({\cal E}_m(\ketbra{\phi_{k+}}{\phi_{k+}}))   \nonumber\\                      
   &&\hspace{0cm}=\sum_k p_k (a_k^2+d_k^2) \nonumber\\                      
   &&\hspace{0.8cm} \times H_2\Bigg(\frac{1}{2}\bigg[1+\sqrt{1-4\eta(1-\eta) \bigg(\frac{d_k^2}{a_k^2+d_k^2}\bigg)^2}\Bigg]\Bigg),
   \label{eq:lowerbound-outputentropy-a}
\end{eqnarray}
where we have used the fact that 
the states ${\cal E}_m(\ketbra{\phi_{k\pm}}{\phi_{k\pm}})$ have the 
same entropy (see equation (\ref{eq:eigenvalues-output-purestateE2})).
% which is equal to
%$H_2\Big(1/2\{1+\big[1-4\eta(1-\eta)d_k^4/(a_k^2+d_k^2)^2\big]^{1/2}\Big)$.
In order to assert that the new ensemble 
$\{\tilde{p}_k,\ket{\tilde{\psi}_k}\}$ 
produces a greater Holevo quantity (\ref{eq:chi-pure}) than the one produced by
$\{{p}_k,\ket{{\psi}_k}\}$ we have to prove that
\begin{eqnarray}
   &&\hspace{0cm}\sum_k p_k  
         H_2\Bigg(\frac{1}{2}\Bigg[1+\sqrt{1-4(1-\eta) d_k^2(2b_k^2 +\eta d_k^2})\Bigg]\Bigg) \ge \nonumber \\
   &&\hspace{0cm}\sum_k p_k (a_k^2+d_k^2) 
         H_2\Bigg(\frac{1}{2}\Bigg[1+\sqrt{1-4\eta(1-\eta) \frac{d_k^4}{(a_k^2+d_k^2)^2}}\Bigg]\Bigg),\nonumber \\
   \label{eq:lowerbound-outputentropy-b}
\end{eqnarray}
the left hand side of (\ref{eq:lowerbound-outputentropy-b}) 
being the last term in (\ref{eq:optimal-chi}).
A sufficient condition for the validity of inequality 
(\ref{eq:lowerbound-outputentropy-b}) is that the
inequality
\begin{eqnarray}
   &&\hspace{0cm}  
         H_2\Bigg(\frac{1}{2}\Bigg[1+\sqrt{1-4(1-\eta)d_k^2 (2b_k^2 +\eta d_k^2})\Bigg]\Bigg) \ge \nonumber \\
   &&\hspace{0cm}(a_k^2+d_k^2) 
         H_2\Bigg(\frac{1}{2}\Bigg[1+\sqrt{1-4\eta(1-\eta) \frac{d_k^4}{(a_k^2+d_k^2)^2}}\Bigg]\Bigg) \nonumber \\
   \label{eq:lowerbound-outputentropy-c}
\end{eqnarray}
holds true for any admissible value of $a_k,\,b_k,\,d_k$, and $\eta$. We checked it 
numerically and it turns
out that this inequality holds; moreover it is tight except for
$\eta=1$, or $b=0$, or $d=0$.   
  
By summarizing the above results, we can state 
that for any ensemble $\{p_k,\ket{\psi_k}\}$, we can find a new one 
$\{\tilde{p}_k,\ket{\tilde{\psi}_k}\}$ of the form (\ref{eq:states-2}), 
whose Holevo quantity is at least as great. 
For this new ensemble the output entropy is given by 
(\ref{eq:optimal-output-entropy}),
whereas the average output entropy is given by (\ref{eq:lowerbound-outputentropy-a}).
%For any $d\neq 0$ the two ensembles are different,
%and moreover, the Holevo quantity relative to the ensemble (\ref{eq:states-1}) \textit{is strictly
%greater} that the one of (\ref{eq:optimal-ensemble-inputstate}-\ref{eq:optimal-ensemble-density-operator}).

\begin{figure}[t!]
  \begin{center}
  \includegraphics[width=7.5cm]{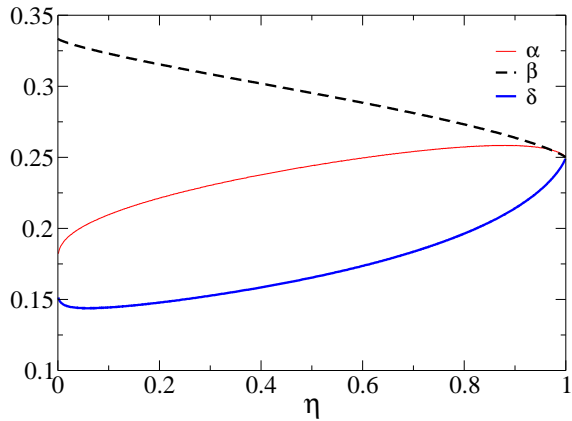}
  \end{center}
  \caption{(Color online) Coefficients $\alpha$ (thin red curve), 
$\beta$ (dashed curve), and
           $\delta$ (thick blue curve) maximizing the Holevo quantity, 
plotted as functions of $\eta$. 
Such coefficients are obtained by numerically solving the 
           optimization problem (\ref{eq:E2-C1-capacity}).}
  \label{fig:Optimal-Bcoefficients}
\end{figure}

We can now find an upper bound to the Holevo quantity 
of ensemble (\ref{eq:states-2}) by considering its average output entropy 
(\ref{eq:lowerbound-outputentropy-a}), and
by taking advantage of the convexity of 
the function $H_2(\frac{1}{2}(1+\sqrt{1-x^2}))$~\cite{giovannetti} with respect to $x$:
\scriptsize
\begin{eqnarray}
   &&\hspace{0cm}\sum_k p_k (a_k^2+d_k^2) 
         H_2\Bigg(\frac{1}{2}\bigg[1+\sqrt{1-4\eta(1-\eta) \bigg(\frac{d_k^2}{a_k^2+d_k^2}\bigg)^2}\Bigg]\Bigg)\,\ge\nonumber\\
   &&\hspace{0.2cm}(\alpha+\delta) 
         H_2\Bigg(\frac{1}{2}\bigg[1+\sqrt{1-4\eta(1-\eta) \bigg(\sum_k p_k \frac{a_k^2+d_k^2}{\alpha+\delta}\frac{d_k^2}{a_k^2+d_k^2}\bigg)^2}\Bigg]\Bigg)=\nonumber\\
&&\hspace{0.2cm}(\alpha+\delta) 
         H_2\Bigg(\frac{1}{2}\bigg[1+\sqrt{1-4\eta(1-\eta) \bigg(\frac{\delta}{\alpha+\delta}\bigg)^2}\Bigg]\Bigg).
   \label{eq:lowerbound-outputentropy-d}
\end{eqnarray}
\normalsize
%where we used that $\delta=\sum_k p_k d_k^2$.
The Holevo 
quantity of the ensemble (\ref{eq:states-2}) is thus upper bounded by 
%of ${\cal E}_m$ that 
\begin{eqnarray}
  &&\hspace{0.0cm}\chi^{*}= \max_{\alpha, \beta, \delta}\Bigg\{-[\alpha +(1-\eta)\delta]\log_2[\alpha +(1-\eta)\delta]+\nonumber\\
  &&\hspace{1.9cm}-2\beta\log_2(\beta)-\,\eta\delta\log_2(\eta\delta)+\nonumber\\
  &&\hspace{0.5cm}      
   -(\alpha+\delta)H_2\Bigg(\frac{1}{2}\bigg[1+\sqrt{1-4\eta(1-\eta) \bigg(\frac{\delta}{\alpha+\delta}\bigg)^2}\Bigg]\Bigg)\Bigg\},\nonumber\\
   \label{eq:E2-C1-capacity}
\end{eqnarray}
%with $\alpha, \beta, \delta$ real and satisfying the constraint $\alpha+2\beta+\delta=1$.
This is precisely the Holevo quantity achievable by
ensemble ${\cal B}$, Sec.~\ref{section:lower-bound}  
equations (\ref{eq:chi-B}) and  (\ref{eq:lowerbound2}),
therefore
%allows us to reach the bound (\ref{eq:E2-C1-capacity}) 
we conclude that (\ref{eq:E2-C1-capacity}) gives 
the $C_1$ classical capacity of  ${\cal E}_m$. 
%(the value of the 
%angle $\theta$ can be any real number).
In Fig. \ref{fig:Optimal-Bcoefficients} we plot the values of the coefficients
$\alpha, \beta, \delta$ which give the maximum of the Holevo quantity for
ensemble ${\cal B}$, whereas the plot of $C_1$ as a function
of $\eta$ is just given by the thick curve of Fig. \ref{fig:bounds}.

It is worth noting that
for $\eta=0$, the maximization problem (\ref{eq:E2-C1-capacity}) does not
admit a unique solution for the coefficients $\alpha$ and $\delta$;
indeed, in this case, the channel deterministically transforms 
$\ket{11}$ into $\ket{00}$,
so that any state 
$\sqrt{\alpha/(\alpha+\delta)}\ket{00}+\sqrt{\delta/(\alpha+\delta)}\ket{11}$
is mapped into $\ket{00}$. To obtain the maximum 
of the Holevo quantity (which
in this case equals $\log_2 3$), we can
arbitrarily choose $\alpha$ and $\delta$, provided that $\alpha+\delta=1/3$.
For a noiseless channel ($\eta=1$), Fig. \ref{fig:Optimal-Bcoefficients}
shows that the optimal coefficients are $\alpha=\beta=\delta=\frac{1}{4}$;
it means that ensemble ${\cal B}$ reduces 
to four orthogonal states, one pair inside the subspace  
$\textrm{span}\{\ket{01},\ket{10}\}$, the other inside the subspace 
 $\textrm{span}\{\ket{00},\ket{11}\}$, each state occurring with equal probability $\frac{1}{4}.$  

\subsection{Is entanglement necessary to achieve $C_1$?}

It is worth to note that the ensemble $\cal B$, allowing
to reach $C_1$, contains entangled states in 
the subspace $\{\ket{00},\ket{11}\}$.
This raises the following question: 
``Is entanglement a necessary ingredient 
to achieve the channel capacity $C_1$?"
In appendix \ref{appx:ent} we show that the answer is positive.
In particular, we show that for any $0<\eta<1$, only the use of entangled 
states allows to achieve $C_1$ and the optimal ensemble is of the form

\begin{eqnarray}
&&\hspace{0.1cm}
   \left\{
   \begin{array}{ll}
    p_{\pm}=\frac{\alpha+\delta}{2}, &
               \ket{\phi_{\pm}}=\sqrt{\frac{\alpha}{\alpha+\delta}}\ket{00}\pm
               \sqrt{\frac{\delta}{\alpha+\delta}}\ket{11},\nonumber\\ %\label{eq:states-1}
               \quad & \quad \nonumber\\
               p_{0}=\beta, &
               \ket{\varphi_{0}}=\ket{01},\nonumber\\
               \quad & \quad \nonumber\\
               p_{1}=\beta, &
               \ket{\varphi_{1}}=\ket{10}.
               \end{array} \right. \\
               \label{eq:ent-states}
\end{eqnarray}

\begin{figure}[t!]
  \begin{center}
  \includegraphics[width=7.5cm]{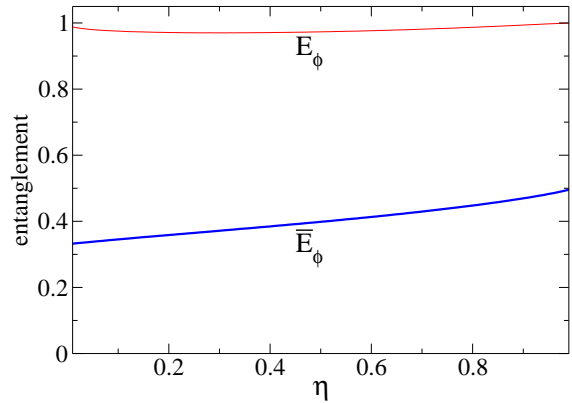}
  \end{center}
  \caption{(Color online) Entanglement $E_\phi$ (thin red curve) 
   of the pure states $\ket{\phi_\pm}$ in the ensemble ${\cal B}$,
   and average entanglement $\overline{E}_\phi=(\alpha+\delta)E_\phi$ (thick 
blue curve) as a function of the transmissivity $\eta$.
   The values of $\alpha,\beta,\delta$ are those ones solving the
   maximization problem (\ref{eq:E2-C1-capacity}).}
  \label{fig:Optimal-Bcoefficients-entanglement}
\end{figure}

One can ask how much entanglement is needed in order to achieve this bound. 
We can answer this question for the above ensemble. It is clear
that we really need entanglement only inside the subspace spanned by 
$\{\ket{00},\,\ket{11}\}$. In Fig. 
\ref{fig:Optimal-Bcoefficients-entanglement} 
we plot the entropy of entanglement
$E_\phi$, defined as the von Neumann entropy of one of the two
reduced states, obtained after tracing over one of the two qubits:
$E_\phi=S(\rho_1)=S(\rho_2)$, with 
$\rho_{1(2)}={\rm Tr}_{2(1)}(\ket{\phi_{\pm}}\bra{\phi_{\pm}})$ 
%and $\rho_2={\rm Tr}_1(\ket{\phi_{\pm}}\bra{\phi_{\pm}})$.
$E_\phi$ quantifies the entanglement content of the states $\ket{\phi_{\pm}}$  
in the ensemble ${\cal B}$, see (\ref{eq:best-ensemble}).
The average entanglement required is given by 
$\overline{E}_\phi=(\alpha+\delta)E_\phi$, since we really 
need entanglement only when we use a state inside the subspace 
spanned by $\{\ket{00},\ket{11}\}$,
which happens with probability $\alpha+\delta$.

\subsection{An explicit formula for $C_1$}
The form of the ensemble (\ref{eq:states-2}) which allows us to maximize
the Holevo quantity of channel ${\cal E}_m$ % giving the $C_1$ classical capacity,
tells us that we can view our memory channel 
as composed of two distinct and parallel
channels acting on two orthogonal subspaces of the four-dimensional
Hilbert space of the two-qubit system: a noiseless channel inside the subspace 
spanned by $\{\ket{10},\ket{01}\}$ and a memoryless amplitude damping channel 
inside the subspace spanned by $\{\ket{00},\ket{11}\}$.
We denote these two channels as ${\cal E}_\varphi$ and
${\cal E}_\phi$, respectively.
In other words, we have proved that, for the fully correlated amplitude 
damping channel ${\cal E}_m$,
the channel capacity $C_1$ is obtained without involving
any coherent superposition of
states from these two different subspaces.
This allows to analytically carry out the optimization (\ref{eq:E2-C1-capacity}). Indeed we can write
\begin{eqnarray}
  &&\hspace{-0.7cm}C_1({\cal E}_m)=\max_{\{p_k,\rho_k\}}{\chi\big({\cal E}_m,\{p_k,\rho_k\}\big)}\,=\nonumber\\
%  &&\hspace{-0.5cm}\max_{\{p_k,\rho_k\}=\{p_{\phi k},\rho_{\phi k}\}\cup\{p_{\varphi k},\rho_{\varphi k}\}}
  &&\hspace{-0.5cm}\max_{\{p_{\phi k},\rho_{\phi k}\}\cup\{p_{\varphi k},\rho_{\varphi k}\}}
{\chi\big({\cal E}_m,\{p_k,\rho_k\}\big)},
    \label{eq:new-argument-01}
\end{eqnarray}
where $\rho_{\phi k}$ is a generic state inside the subspace 
$\{\ket{00},\ket{11}\}$,
whereas $\rho_{\varphi k}$ is a generic state inside the subspace $\{\ket{01},\ket{10}\}$.
Now we call $p=\sum_k p_{\phi k}$, 
and consequently we have that $\sum_k p_{\varphi k}=1-p$.
We can then write
\begin{eqnarray}
  \hspace{-2cm} \rho &=& \sum_k p_k \rho_k \,=\, 
    \sum_k p_{\phi k}\, \rho_{\phi k}\,+\, \sum_k p_{\varphi k}\, \rho_{\varphi k}\nonumber\\
  &&= p \sum_k \frac{p_{\phi k}}{\sum_{k'} {p}_{\phi {k'}}} \rho_{\phi k}\,+\, 
      (1-p)\sum_k \frac{p_{\varphi k}}{\sum_{k'} {p}_{\varphi {k'}}} \rho_{\varphi {k'}}\nonumber\\
  &&= p \sum_k \tilde{p}_{\phi k}\, \rho_{\phi k}\,+\, 
      (1-p)\sum_k \tilde{p}_{\varphi k}\, \rho_{\varphi k}\nonumber\\
  &&= p \,\rho_{\phi}\,+\, (1-p)\,\rho_{\varphi},
    \label{eq:new-argument-02}
\end{eqnarray}
where we have set
\begin{eqnarray}
&& \tilde{p}_{\phi k}\equiv\frac{p_{\phi k}}{\sum_{k'} {p}_{\phi {k'}}},\quad
\rho_{\phi}\equiv\sum_k \tilde{p}_{\phi k}\,\rho_{\phi k},\nonumber\\ 
&& \tilde{p}_{\varphi k}\equiv\frac{p_{\varphi k}}{\sum_{k'} {p}_{\varphi {k'}}},\quad 
 \rho_{\varphi}\equiv\sum_k \tilde{p}_{\varphi k}\,\rho_{\varphi k}.
    \label{eq:new-argument-03}
\end{eqnarray}
Note that $\textrm{Tr}[\rho_\phi]=\textrm{Tr}[\rho_\varphi]=1$. 
The first term of the Holevo quantity (\ref{eq:new-argument-01}) is given by
\begin{eqnarray}
 && \hspace{0.2cm}S\Big({\cal E}_m\Big(\sum_k p_k \rho_k\Big)\Big)\,=S\Big(p\,{\cal E}_m (\rho_\phi)\,+\,(1-p)\,
{\cal E}_m(\rho_\varphi)\Big)\,\nonumber\\
 &&\hspace{1cm}=\, H_2(p)\,+\,p\,S\big({\cal E}_m (\rho_\phi)\big)\,+
 \,(1-p)\,S\big({\cal E}_m(\rho_\varphi)\big)\nonumber\\
 &&\hspace{1cm}=\, H_2(p)\,+\,p\,S\big({\cal E}_m (\sum_k \tilde{p}_{\phi k}\,\rho_{\phi k})\big)\,+\nonumber\\
 &&\hspace{2.5cm}+\,(1-p)\,S\big({\cal E}_m(\sum_k \tilde{p}_{\varphi k}\,\rho_{\varphi k})\big)\;,
    \label{eq:new-argument-05}
\end{eqnarray}
%where we have used the linearity of quantum operations 
where the second equality is due to the fact that
 the two output states in the above equation are 
supported on orthogonal subspaces and can therefore 
be independently and simultaneously diagonalized.
Now we turn to the second term of the Holevo quantity, namely
\begin{eqnarray}
 && \hspace{-0.2cm}\sum_k p_k S\big({\cal E}_m (\rho_k)\big) \,=\, \nonumber\\
 && \hspace{0.2cm}= \sum_k p_{\phi k} S\big({\cal E}_\phi (\rho_{\phi k})\big)\,+\,
    \sum_k p_{\varphi k} S\big({\cal E}_\varphi (\rho_{\varphi k})\big)\nonumber\\
 && \hspace{0.2cm}= p \sum_k \tilde{p}_{\phi k} S\big({\cal E}_\phi (\rho_{\phi k})\big)\,+\,
   (1-p) \sum_k \tilde{p}_{\varphi k} S\big({\cal E}_\varphi (\rho_{\varphi k})\big),\nonumber\\
    \label{eq:new-argument-06}
\end{eqnarray}
where we have used the quantities defined in (\ref{eq:new-argument-03}).
From (\ref{eq:new-argument-05}) and (\ref{eq:new-argument-06}) we obtain
\begin{eqnarray}
  &&\hspace{0cm}{\chi\big({\cal E}_m,\{p_k,\rho_k\}\big)}\,=\,%S\Big({\cal E}_m\Big(\sum_k p_k \rho_k\Big)\Big)-
  %\sum_k p_k S\big({\cal E}_m (\rho_k)\big)
  \nonumber\\
    %    \hspace{-1cm}\max_{\{p_k,\rho_k\}=\{p_{\phi k},\rho_{\phi k}\}\cup\{p_{\varphi k},\rho_{\varphi k}\}}
   %\chi\big({\cal E}_\phi\otimes{\cal E}_\varphi,\{p_k,\rho_k\}\big)
%  &&\hspace{0.3cm}=H_2(p)+p\,S\big({\cal E}_\phi (\sum_k \tilde{p}_{\phi k}\,\rho_{\phi k})\big)\,+\nonumber\\
%     &&\hspace{0.7cm} 
%    +\,(1-p)S\big({\cal E}_\varphi(\sum_k \tilde{p}_{\varphi k}\,\rho_{\varphi k})\big)
%   \,-\,p \sum_k \tilde{p}_{\phi k} S\big({\cal E}_\phi (\rho_{\phi k})\big)
%  +\nonumber\\
% &&\hspace{0.7cm} \,-
%   (1-p) \sum_k \tilde{p}_{\varphi k} S\big({\cal E}_\varphi (\rho_{\varphi k})\big)\nonumber\\
% &&\hspace{0.3cm}=H_2(p)\,+p\,\chi\big({\cal E}_\phi,\{\tilde{p}_{\phi k},\rho_{\phi k}\}\big)\,+\nonumber\\
% &&\hspace{2.1cm}   +\,(1-p)\,\chi\big({\cal E}_\varphi,\{\tilde{p}_{\varphi k},\rho_{\varphi k}\}\big)\nonumber\\
 &&\hspace{0.3cm}=H_2(p)+p\,\chi_\phi\big(\{\tilde{p}_{\phi k},\rho_{\phi k}\}\big)
    +\,(1-p)\,\chi_\varphi\big(\{\tilde{p}_{\varphi k},\rho_{\varphi k}\}\big),\nonumber\\
    \label{eq:new-argument-07}
\end{eqnarray}
where we have defined 
\begin{eqnarray}
&&\chi_\phi\big(\{\tilde{p}_{\phi k},\rho_{\phi k}\}\big)\equiv
 \chi\big({\cal E}_\phi,\{\tilde{p}_{\phi k},\rho_{\phi k}\}\big),
\nonumber\\ 
&&\chi_\varphi\big(\{\tilde{p}_{\varphi k},\rho_{\varphi k}\}\big)\equiv
 \chi\big({\cal E}_\varphi,\{\tilde{p}_{\varphi k},\rho_{\varphi k}\}\big).
\end{eqnarray}
The maximization problem (\ref{eq:new-argument-01}) 
is therefore equivalent to
\begin{eqnarray}
  &&\hspace{-0cm}C_1({\cal E}_m)=\max_{\{p_k,\rho_k\}}{\chi\big({\cal E}_m,\{p_k,\rho_k\}\big)}\nonumber\\
  &&\hspace{0.1cm}
  =\max_{\{{p}_{\phi k},\rho_{\phi k}\},\{{p}_{\varphi k},\rho_{\varphi k}\}}
  \Big[H_2(p)\,+\,p\,
           \chi_\phi\big(\{\tilde{p}_{\phi k},\rho_{\phi k}\}\big)\nonumber\\
  &&\hspace{4cm}    +\,(1-p)\,
          \chi_\varphi\big(\{\tilde{p}_{\varphi k},\rho_{\varphi k}\}\big)\Big]\nonumber\\
  &&\hspace{0.1cm}
  =\max_{p\in[0,1]}\Big[H_2(p)\,+\,p\,
      \max_{\{\tilde{p}_{\phi k},\rho_{\phi k}\}}
           \chi_\phi\big(\{\tilde{p}_{\phi k},\rho_{\phi k}\}\big)\nonumber\\
  &&\hspace{2.6cm}    +\,(1-p)\,
      \max_{\{\tilde{p}_{\varphi k},\rho_{\varphi k}\}}
          \chi_\varphi\big(\{\tilde{p}_{\varphi k},\rho_{\varphi k}\}\big)\Big]\nonumber\\
  &&\hspace{0.1cm}
  =\max_{p\in[0,1]}\Big[H_2(p)\,+\,p\,C_{\phi_1}\,+\,(1-p)\,C_{\varphi_1}\Big],
%{\{p_k,\rho_k\}=\{p_{\phi k},\rho_{\phi k}\}\cup\{p_{\varphi k},\rho_{\varphi k}\}},
%{\chi\big({\cal E}_\phi\otimes{\cal E}_\varphi,\{p_k,\rho_k\}\big)},
    \label{eq:new-argument-08}
\end{eqnarray}
where $C_{\phi_1}$ and $C_{\varphi_1}$ are respectively the classical
product state capacity 
\begin{eqnarray}
  &&  C_{\phi_1}\,=\,\max_{\{\tilde{p}_{\phi k},\rho_{\phi k}\}}
        \chi\big(\{\tilde{p}_{\phi k},\rho_{\phi k}\}\big),  \label{eq:new-argument-09}\\
  &&  C_{\varphi_1}\,=\,\max_{\{\tilde{p}_{\varphi k},\rho_{\varphi k}\}}
        \chi\big({\cal E}_m,\{\tilde{p}_{\varphi k},\rho_{\varphi k}\}\big).
\end{eqnarray}
The maximization (\ref{eq:new-argument-08}) over $p$ can then 
be simply achieved by studying
the first derivative of $G(p)\equiv H_2(p)\,+\,p\,C_{\phi_1}\,+\,(1-p)\,C_{\varphi_1}$ 
with respect to $p$: 
\begin{eqnarray}
%&&\hspace{-1cm}\frac{\partial}{\partial p}\big[H_2(p)\,+\,p\,C_{\phi_1}\,+\,(1-p)\,C_{\varphi_1}\big]\,=\,\nonumber\\
&&\hspace{-1cm}\frac{\partial G(p)}{\partial p}\,=\,
\log_2\frac{1-p}{p}\,+\,C_{\phi_1}\,-\,C_{\varphi_1}.
  \label{eq:new-argument-10}
\end{eqnarray}
A maximum is found for:
\begin{eqnarray}
  p_{opt}\,=\,\frac{1}{1+2^{C_{\varphi_1}-C_{\phi_1}}}\,=\,\frac{1}{1+2^{1-C_{ad,1}}},
  \label{eq:new-argument-11}
\end{eqnarray}
since $C_{\varphi_1}=1$ and $C_{\phi_1}$ is the product state capacity 
$C_{ad,1}$  of the memoryless amplitude 
damping channel (\ref{eq:ampl-damping-channel}) \cite{giovannetti},
which is given by
\begin{eqnarray}
&& \hspace{-0.5cm} C_{ad,1}\,=\max_{p_1 \in[0,1]}\Bigg[H_2(\eta p_1)\,+\,\nonumber\\
&&  \hspace{2cm}      -\,H_2\Big(\frac{1}{2}\Big(1+\sqrt{1-4\eta(1-\eta)p_1^2}\Big)\Big)\Bigg].
  \label{eq:new-argument-12}
\end{eqnarray}
It is worth noting that the optimal value of $p_1$ in 
(\ref{eq:new-argument-12})
also gives the population of 
the single qubit state $\ket{1}$, in the
density operator describing the ensemble which maximizes the
single-use (and single-qubit) Holevo quantity for the memoryless amplitude 
damping channel \cite{giovannetti}. 

\begin{figure}[h!]
  \begin{center}
  \includegraphics[width=7.5cm]{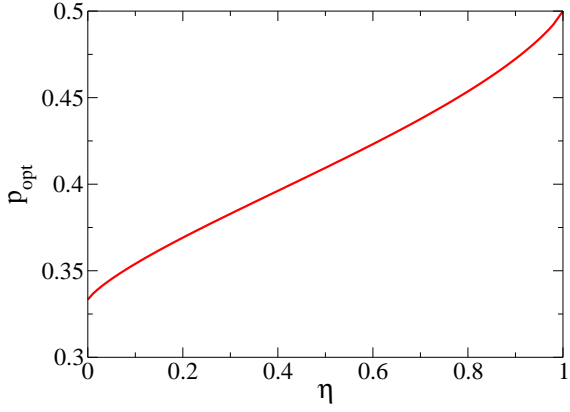}
  \end{center}
  \caption{Plot of $p_{opt}$ (Eq. (\ref{eq:new-argument-11})) 
as functions of $\eta$.}
  \label{fig:p_opt}
\end{figure}

We can conclude that the $C_1$ capacity of the memory channel ${\cal E}_m$ is
\begin{eqnarray}
  C_1({\cal E}_m)\,=\,1\,+\,H_2(p_{opt})\,-\,p_{opt}\,(1-C_{ad,1}).
  \label{eq:new-argument-13}
\end{eqnarray}
%($p_{opt}$ is given by (\ref{eq:new-argument-11}))
Equation (\ref{eq:new-argument-13})
provides an explicit solution to (\ref{eq:E2-C1-capacity}), once $C_{ad,1}$
is known.
In Fig. \ref{fig:p_opt} we show the optimal value $p_{opt}$ 
as a function of the channel transmissivity $\eta$.
Note that the value of $p_{opt}$ tells us the weight of the subspace
spanned by  
$\{\ket{00},\ket{11}\}$ in achieving the $C_1$ capacity of  
the channel ${\cal E}_m$. Let we consider two limiting cases.
As expected, for $\eta=0$ we have that $C_{\phi_1}=0$ and therefore by 
(\ref{eq:new-argument-11})
we find $p_{opt}=1/3$, while for $\eta=1$ we have that $C_{\phi_1}=1$ 
and $p_{opt}=1/2$.

From the maximization procedure we depicted, it is clear that the 
probability $\delta/(\alpha+\delta)$, which gives the population of the 
state $\ket{11}$ in the density operator describing the optimal ensemble,
%(\ref{eq:best-ensemble}), 
normalized by the probability that a state picked up from this ensemble belongs
to the subspace spanned by $\{\ket{00},\ket{11}\}$,
is the same of the optimal $p_1$ ensuring the achievement of $C_{ad,1}$ in
 (\ref{eq:new-argument-12}) (see equation (\ref{eq:new-argument-09})).

\section{Quantum Capacity}
\label{sec:quantum-capacity}

The quantum capacity $Q$ concerns the channel ability to convey quantum information.
It can be computed as~\cite{lloyd,barnum,devetak}
\begin{equation}
Q \,=\, \lim_{n\to\infty} \frac{Q_n}{n},
\quad \quad
Q_n\,=\,\max_{\rho^{(n)}} I_c(\mathcal{E}_{m}^{\otimes n},\rho^{(n)}),
\label{qinfo}
\end{equation}
where $\rho^{(n)}$ is an input state for $n$ channel uses and 
\begin{equation}
I_c(\mathcal{E}_{m}^{\otimes n},\rho^{(n)})\,=\,
S\big(\mathcal{E}_{m}^{\otimes n}\big( \rho^{(n)} \big) \big)\,-\,
S_e(\mathcal{E}_{m}^{\otimes n},\rho^{(n)})
\label{coinfo}
\end{equation}
is the \textit{coherent information}~\cite{schumachernielsen}.
$S_e(\mathcal{E}_{m}^{\otimes n},\rho^{(n)})$ is
the \textit{entropy exchange}~\cite{schumacher}, defined as 
\begin{equation}
S_e(\mathcal{E}_{m}^{\otimes n},\rho^{(n)})
=S\big[\big(\mathcal{I}\otimes \mathcal{E}_{m}^{\otimes n}\big)
\big(|\Psi\rangle\langle\Psi\big)\big],
\end{equation}
where $|\Psi\rangle$ is any purification of $\rho^{(n)}$.
That is, we consider the system ${\textsf S}$, described by 
the density operator $\rho^{(n)}$, as a part of a larger quantum
system ${\textsf R}{\textsf S}$;
$\rho=\mathrm{Tr}_{\textsf R} |\Psi\rangle\langle\Psi|$ and
the reference system
${\textsf R}$ evolves trivially, according to the identity
superoperator $\mathcal{I}$.
Note that maximization (\ref{qinfo}) has to be carried with respect to a generic
density operator $\rho^{(n)}$ belonging to the Hilbert space relative
to $n$ uses of the channel $\mathcal{E}_{m}$ 
(described by the superoperator $\mathcal{E}_{m}^{\otimes n}$).

\subsection{Quantum capacity for channel transmissivity 
$\frac{1}{2}\le \eta\le 1$}

In order to proceed to the calculation of the quantum capacity 
we will use the fact that the channel ${\cal E}_m$ is degradable 
\cite{degradable} for
$\frac{1}{2}\le \eta\le 1$, as shown in Appendix \ref{appx:degradability}.
Degradability implies that regularization (\ref{qinfo})
is no longer necessary, i.e., the quantum capacity is given by the
``single-letter'' formula, $Q=Q_1$:
\begin{equation}
  Q({\cal E}_m)\,=\,\max_{\rho}\,I_c({\cal E}_m,\rho), 
\quad \eta\,\in\,[\frac{1}{2},1],
\label{eq:QuantumCapacity}
\end{equation}
where $\rho$ belongs to the Hilbert space relative to a single use of channel ${\cal E}_m$.

%\subsection{Maximization of the Coherent Information}
\label{sec:Maximization of the Coherent Information}

%First we discuss the quantum capacity of channel ${\cal E}_m$ for
%$\eta\ge 1/2$. 
The coherent information is given by:
\begin{eqnarray}
  I_c({\cal E}_m,\rho)\,&=&\,S({\cal E}_m(\rho))\,-S_e({\cal E}_m,\rho)\nonumber\\
                          &=& S(\rho')\,-\,S(\rho^{\textsf{E}'}),
  \label{eq:CoherentInformation}
\end{eqnarray}
where
$S_e({\cal E}_m,\rho)\,=\,S(\rho^{\textsf{E}'})$ 
is the entropy exchange related to the channel~\cite{schumachernielsen}.
Here $\rho$ is a generic input state for the channel ${\cal E}_m$,
$\rho'={\cal E}_m(\rho^{\textsf{S}})$ and $\rho^{\textsf{E}'}$ are given by 
(\ref{eq:output-genericstate}) and (\ref{eq:ancilla-output-genericstate}),
being $\textsf{E}$ a fictitious environment allowing for a unitary 
representation of the map ${\cal E}_m$ (see Appendix \ref{appx:degradability}). 

Our target is to find the class of input states which allow to
solve problem (\ref{eq:QuantumCapacity}),
i.e. to maximize the coherent information (\ref{eq:CoherentInformation}).
To this end, we first notice that for any two-qubit density operator $\rho$, we can build 
a diagonal density operator as follows
\begin{eqnarray}
&&\tilde{\rho}=\frac{1}{4}(\rho+\sum_{i=1}^3{\cal R}_i\,\rho{\cal R}_i)=%\nonumber\\
%&&\hspace{0.5cm}=
\left(\begin{array}{cccc}
  \alpha & 0 & 0 & 0 \\
   0 & \beta & 0 & 0 \\
   0 & 0 & \gamma & 0 \\
   0 & 0 & 0 & \delta \\
  \end{array}\right), \label{eq:rhotilde}
\end{eqnarray}
whose coherent information is at least as large as 
the one 
related to ${\rho}$:
%(\ref{eq:rhotilde})
\begin{eqnarray}
 && \hspace{-0.0cm}I_c({\cal E}_m,\tilde{\rho})\,=\,I_c\Big({\cal E}_m,\frac{1}{4}
    \Big(\rho\,+\sum_{i=1}^3\,{\cal R}_i\rho{\cal R}_i\Big)\Big)\nonumber\\
 && \hspace{0.6cm}\ge \frac{1}{4}I_c\big({\cal E}_m,\rho)  \,+\,\frac{1}{4}\sum_{i=1}^3
    I_c\big({\cal E}_m,{\cal R}_i\rho{\cal R}_i\big)\,=\nonumber\\
 && \hspace{0.6cm}= \frac{1}{4}I_c\big({\cal E}_m,\rho)  
    \,+\,\frac{1}{4}\sum_{i=1}^3 S({\cal E}_m({\cal R}_i\rho{\cal R}_i))\,+\nonumber\\
 && \hspace{2.7cm}    \,-\,\frac{1}{4}\sum_{i=1}^3 S_e({\cal E}_m,{\cal R}_i\rho{\cal R}_i)\,=\nonumber\\  
&& \hspace{0.6cm}\,=\,I_c({\cal E}_m,\rho).
\end{eqnarray}
Here, the inequality derives from the fact that the coherent information
of a degradable channel is a concave function~\cite{wolf2007} and we have 
used the covariance properties of the channel.
Finally, since ${\cal R}_i$ can only change the sign of coherences of 
the input state,
%(in particular of $\varsigma$, see appendix \ref{appx:degradability}), 
the von Neumann
entropy of $\rho^{\textsf{E}'}$ does not change
when we replace  $\rho$ by ${\cal R}_i\rho{\cal R}_i$:
$S_e({\cal E}_m,{\cal R}_i\rho{\cal R}_i)=S_e({\cal E}_m,\rho)$,
as one can see by Eq.~(\ref{eq:ancilla-output-genericstate}).
%So we have demonstrated that:
%\begin{equation}
% I_c({\cal E}_m,\tilde{\rho}^S)\, \ge\, I_c({\cal E}_m,\rho)
%\end{equation}

Now we build a new state from $\tilde{\rho}$: 
\begin{eqnarray}
&&\overline{\rho}=\frac{1}{2}\tilde{\rho}+
         \frac{1}{2}{\cal S}_\textrm{w}\tilde{\rho}{\cal S}_\textrm{w}%\nonumber\\
=\left(\begin{array}{cccc}
  \alpha & 0 & 0 & 0 \\
   0 & \frac{\beta+\gamma}{2} & 0 & 0 \\
   0 & 0 & \frac{\beta+\gamma}{2} & 0 \\
   0 & 0 & 0 & \delta \\
  \end{array}\right).
\end{eqnarray}
This new density operator exhibits a coherent information
greater than or equal to $\tilde{\rho}$, since 
\begin{eqnarray}
 &&I_c({\cal E}_m,\overline{\rho})\,=\,
   I_c({\cal E}_m,\frac{1}{2}\tilde{\rho}+
         \frac{1}{2}{\cal S}_\textrm{w}\tilde{\rho}{\cal S}_\textrm{w})\nonumber\\
&&\hspace{0.2cm}\, \ge\, 
 \frac{1}{2}I_c({\cal E}_m,\tilde{\rho}) \,+\, 
 \frac{1}{2}I_c({\cal E}_m,{\cal S}_\textrm{w}\tilde{\rho}{\cal S}_\textrm{w})=\nonumber\\
&&\hspace{0.2cm}= \frac{1}{2}I_c({\cal E}_m,\tilde{\rho})+
    \frac{1}{2}\Big[S({\cal E}_m({\cal S}_\textrm{w}\tilde{\rho}{\cal S}_\textrm{w}))\,-\,
              S_e({\cal E}_m({\cal S}_\textrm{w}\tilde{\rho}{\cal S}_\textrm{w}))\big]
\nonumber\\
 &&\hspace{0.2cm}=I_c({\cal E}_m,\tilde{\rho}).
\end{eqnarray}
In the above equation we have again exploited the concavity of the coherent 
information 
for degradable channels in getting the inequality;
then we have used the covariance property (\ref{eq:Swap-cov}).
%and the fact that the von Neuman entropy does not change under unitary operation.
% $S({\cal E}_m({\cal S}_\textrm{w}\tilde{\rho}{\cal S}_\textrm{w}))=S({\cal E}_m,\tilde{\rho}S)$
%since the output entropy  does not change 
%(see eq. (\ref{eq:output-entropy}))
%if we exchange $\beta$ with $\gamma$, whereas 
For the entropy exchange we have 
$S_e({\cal E}_m({\cal S}_\textrm{w}\tilde{\rho}{\cal S}_\textrm{w}))=S_e({\cal E}_m(\tilde{\rho}))$
since it does not depend on $\beta$ and $\gamma$ 
(see equation (\ref{eq:ancilla-output-genericstate})).
%, so we have that
%$I_c({\cal E}_m,\tilde{\rho}^S)=I_c({\cal E}_m,\tilde{\rho}^S(\beta\leftrightarrow\gamma))$.

\begin{figure}[t!]
  \begin{center}
  \includegraphics[width=7.5cm]{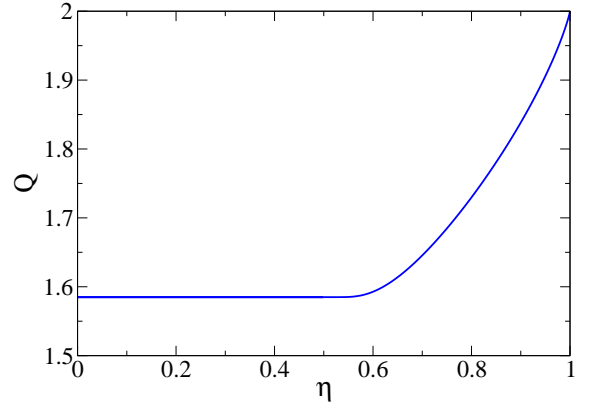}
  \end{center}
  \caption{Plot of the quantum capacity $Q$ of
           ${\cal E}_m$ as a function of $\eta$.
           For $\eta \ge 1/2$, $Q$ is given by the numerical solution 
           of the maximization task (\ref{eq:QuantumCapacity-1})
           (the searching step for $\alpha$, $\beta$, $\delta$ 
           is $10^{-4}$).
           For $\eta<0.5$ the quantum capacity turns out to be constant and
           equal to $\log_2 3$.}
  \label{fig:QCapacity}
\end{figure}

\begin{figure}[t!]
  \begin{center}
  \includegraphics[width=7.5cm]{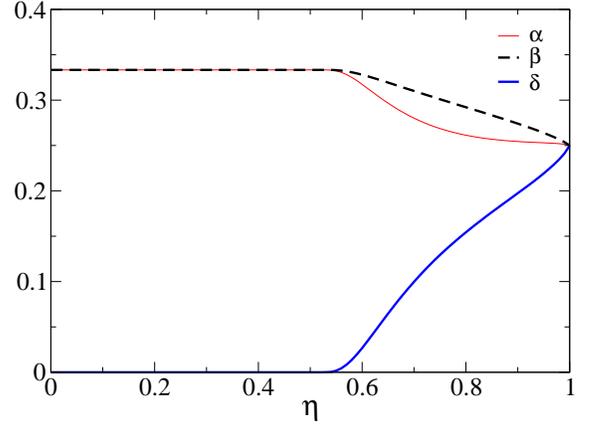}
  \end{center}
  \caption{Plot of the coefficients $\alpha$ (thin curve), $\beta$ (dashed curve) and
           $\delta$ (thick curve) which (numerically) solve the optimization problem (\ref{eq:QuantumCapacity-1}),
           as function of $\eta$.}
  \label{fig:Optimal-QCcoefficients}
\end{figure}

We conclude that the quantum capacity (\ref{eq:QuantumCapacity}) can be
derived by maximizing the coherent information with respect the diagonal state 
\begin{equation}
\overline{\rho}=
\left(\begin{array}{cccc}
  \alpha & 0 & 0 & 0  \\
   0 & \beta & 0 & 0  \\
   0 & 0 & \beta & 0  \\
   0 & 0 & 0 & \delta \\
  \end{array}\right),
\label{eq:rhobar}
\end{equation}
since we have demonstrated that for each $\rho$ we can 	
construct a density operator $\bar{\rho}$ of the form (\ref{eq:rhobar})
 whose coherent information is at least as 
great.
Therefore, for $\eta\ge \frac{1}{2}$ the quantum capacity is given by
\begin{eqnarray}
  Q({\cal E}_m)\,&=&\,\max_{\overline{\rho}^S}\,I_c({\cal E}_m,\overline{\rho}^S)\,=\,\nonumber\\
 &=&\max_{\overline{\rho}^S}\,\Big\{S\big({\cal E}_m\big(\overline{\rho}^S\big)\big)
     \,-\,S_e\big({\cal E}_m,\overline{\rho}^S\big)\Big\}\nonumber\\
&=&\,\max_{\alpha,\beta,\delta}\Big\{
   -[\alpha+(1-\eta)\delta]\log_2[\alpha+(1-\eta)\delta]\,+\nonumber\\
&&\hspace{1.0cm}-\,2\beta\log_2\beta\,-\,\eta \delta \log_2\eta\delta\, +\,\nonumber\\
&&\hspace{1cm}    +\,[1-(1-\eta)\delta]\log_2[1-(1-\eta)\delta]\,+\nonumber\\
&&\hspace{1cm}   +\,(1-\eta)\delta\log_2[(1-\eta)\delta]
\Big\},
\label{eq:QuantumCapacity-1}
\end{eqnarray}
with the constraint $\alpha+2\beta+\delta=1$.
In Fig.~\ref{fig:QCapacity} we plot the quantum capacity $Q$ of the channel 
${\cal E}_m$ as a result of the maximization problem 
(\ref{eq:QuantumCapacity-1}),
and in Fig.~\ref{fig:Optimal-QCcoefficients} we report
the relative populations of the input state
(\ref{eq:rhobar}). The results are displayed 
for $\eta\,\in\,[0,1]$, but we stress that (\ref{eq:QuantumCapacity-1})
give us the quantum capacity only for $\eta\,\in\,[\frac{1}{2},1]$.
%in the range $\eta\,\in\,[0,\frac{1}{2}[$ it only provides a lower bound
%for the quantum capacity. 
Notice that the curve reported in Fig. 
\ref{fig:QCapacity} is higher than the one derived in Ref. \cite{Jahangir}, 
where only a particular class of product input states were considered.

\subsection{Quantum capacity for channel transmissivity 
$0\le \eta <\frac{1}{2}$}

For $\eta < 1/2$, %, we  already found an lower bound ($\log_23$), but 
we cannot use the subadditivity argument
provided by degradability in order to find the channel quantum 
capacity. However, we notice that ${\cal E}_{m}$ has the following property
\begin{equation}
   {\cal E}_{m,\eta_2 \eta_1}={\cal E}_{m,\eta_2}\circ{\cal E}_{m,\eta_1},
\label{eq:channel-property}
\end{equation}
where we have used ${\cal E}_{m,x}$ to indicate a channel ${\cal E}_{m}$
with transmissivity $x$.
Now we choose $\eta_1,\eta_2$ such that $\eta_1=1/2$ and $\eta_2 \in [0,1[$,
then $\eta_2\eta_1 \in [0,1/2[$. By considering $n$ channel uses and 
applying the quantum data processing 
inequality~\cite{barnum}
we obtain
\begin{equation}
I_c({\cal E}_{m,\eta_2\eta_1}^{\otimes n},{{\rho}}^{(n)})\,\le\,
   I_c({\cal E}_{m,\frac{1}{2}}^{\otimes n},{{\rho}}^{(n)}),
\label{eq:data-process-ineq}
\end{equation}
since
${\cal E}_{m,\eta_2\eta_1}^{\otimes n}={\cal E}_{m,\eta_2}^{\otimes n}\circ{\cal E}_{m,\eta_1}^{\otimes n}$.
%(${{\rho}}^{(n)}$ is the input density operator for $n$ channel uses).
Hence, for $\eta<1/2$, the quantum capacity is given by
\begin{eqnarray}
 Q({\cal E}_m)\,&=&\,\lim_{n\to \infty}\max_{{{\rho}}^{(n)}}\,
         I_c\big({\cal E}_m^{\otimes n},{{\rho}}^{(n)}\big)\,\nonumber\\
    &\le& \lim_{n\to \infty}\max_{{{\rho}}^{(n)}}\,
         I_c\big({\cal E}_{m,\frac{1}{2}}^{\otimes n},{{\rho}^{S}}^{(n)}\big)\nonumber\\
    &\le&\max_{{\rho}}\,I_c\big({\cal E}_{m,\frac{1}{2}},{\rho}\big)\,=\,\log_2 3,
\label{eq:QuantumCapacity-2}
\end{eqnarray}
where the second inequality holds since for $\eta=1/2$ the channel is degradable,
whereas the last equality is numerically provided by (\ref{eq:QuantumCapacity-1}).
It is easy to prove that $\log_2(3)$ is also a lower bound 
for the channel quantum capacity, since the three-dimensional 
subspace spanned by $\{\ket{00},\,\ket{01},\,\ket{10}\}$ is noiseless.
We can therefore conclude that, for $\eta<1/2$, $Q({\cal E}_m)=\log_23$.

\section{Entanglement-assisted classical capacity}
\label{sec:entass-capacity} 

The entanglement-assisted classical capacity $C_E$
gives the maximum amount of classical information
that can be reliably transmitted 
down the channel per channel use, provided 
the sender and the receiver share infinite prior  
entanglement resources.
%~\cite{bennett-shor,bennett1999}.
It can be computed as~\cite{bennett-shor,bennett1999}
\begin{equation}
C_E \,=\max_{\rho} I(\mathcal{E}_{m},\rho),
\label{CECapacity}
\end{equation}
where the maximization is performed over 
the input state $\rho$ for a single use of the channel
$\mathcal{E}_{m}$ and 
\begin{equation}
I(\mathcal{E}_{m},\rho) \,= S(\rho) + 
I_c (\mathcal{E}_{m},\rho)
\end{equation}
differs from the coherent information $I_c$,
defined in Eq.~(\ref{eq:CoherentInformation}),
by the addition of the input-state entropy $S(\rho)$.
Since $S(\rho)=S(\rho^{\textsf R})$ and the reference
system $\textsf{R}$ evolves trivially, then 
\begin{equation}
I(\mathcal{E}_{m},\rho)
=S(\rho^{\textsf R})+S(\mathcal{E}_{m}(\rho))-
   S[(\mathcal{I}\otimes\mathcal{E}_{m})(\ketbra{\Psi}{\Psi})]
\end{equation}
is the output \textit{quantum 
mutual information}~\cite{nielsen-chuang} between the system 
$\textsf{S}$ and the reference system $\textsf{R}$.
Note that, due to the subadditivity of 
$I$~\cite{adami-Cerf}, no regularization as in 
(\ref{qinfo}) is required
to obtain $C_E$.  

\subsection{Maximization of the quantum mutual information 
$I(\mathcal{E}_{m},\rho)$}
\label{sec:Maximization-channel-MutualInf}

\begin{figure}[t!]
  \begin{center}
  \includegraphics[width=7.5cm]{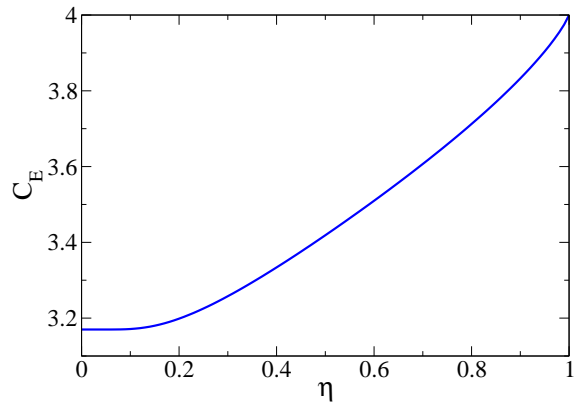}
  \end{center}
  \caption{Plot of the entanglement-assisted capacity $C_E$ of
           ${\cal E}_m$ as a function of $\eta$.
           $C_E$ is obtained from the numerical solution 
           of the maximization task (\ref{eq:CECapacity})
           (the searching step for $\alpha$, $\beta$, $\delta$ 
           is $10^{-4}$).
           For $\eta \to 0$, the entanglement-assisted capacity tends 
to the value $2\log_2 3$.}
  \label{fig:CECapacity}
\end{figure}

\begin{figure}[t!]
  \begin{center}
 \includegraphics[width=7.5cm]{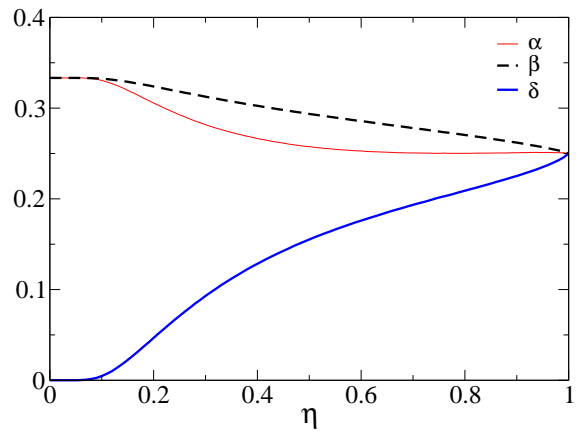}
  \end{center}
  \caption{Plot of the coefficients $\alpha$ (thin curve), $\beta$ (dashed curve) and
           $\delta$ (thick curve) which (numerically) solve the 
optimization problem (\ref{eq:CECapacity}),
           as functions of $\eta$.}
  \label{fig:Optimal-CECcoefficients}
\end{figure}

By following a similar argument as the one exploited in 
deriving Eq.~(\ref{eq:QuantumCapacity-1}) for the quantum 
capacity, we obtain 

\begin{eqnarray}
  C_E({\cal E}_m)\,&=&\,\max_{\overline{\rho}}\,I({\cal E}_m,\overline{\rho})\,=\,\nonumber\\
 &=&\max_{\overline{\rho}}\,\Big\{S\big({\cal E}_m\big(\overline{\rho}\big)\big)
     \,+\,I_c\big({\cal E}_m,\overline{\rho}\big)\Big\}\nonumber\\
&=&\,\max_{\alpha,\beta,\delta}\Big\{-\alpha\log_2\alpha\,-\,\delta\log_2\delta+\nonumber\\
&&\hspace{1.0cm}   -[\alpha+(1-\eta)\delta]\log_2[\alpha+(1-\eta)\delta]\,+\nonumber\\
&&\hspace{1.0cm}-\,4\beta\log_2\beta\,-\,\eta \delta \log_2\eta\delta\, +\,\nonumber\\
&&\hspace{1cm}    +\,[1-(1-\eta)\delta]\log_2[1-(1-\eta)\delta]\,+\nonumber\\
&&\hspace{1cm}   +\,(1-\eta)\delta\log_2[(1-\eta)\delta]
\Big\},
\label{eq:CECapacity}
\end{eqnarray}
where the optimization is over a diagonal input state 
$\overline{\rho}$ of the form (\ref{eq:rhobar})
(with the constraint $\alpha+2\beta+\delta=1$).
We plot the entanglement-assisted classical 
capacity $C_E$ of the channel ${\cal E}_m$ 
as a result of the maximization problem 
(\ref{eq:CECapacity}) in Fig.~\ref{fig:CECapacity},
and the relative populations of the optimal ensemble
in Fig.~\ref{fig:Optimal-CECcoefficients}. 


Note that for $\eta=0$ the entanglement-assisted 
classical capacity is 
$2\log_23$, as it turns out from the optimization
problem (\ref{eq:CECapacity}), see Fig.~\ref{fig:CECapacity}.
Indeed, in this case we have at our disposal a noiseless subspace,
spanned by $\{\ket{00},\ket{01},\ket{10}\}$, of dimension 
$d=3$. This means that we can use
a quantum superdense coding protocol 
(see Ref.~\cite{bennett-shor}) in this subspace,
achieving a transmission rate of $2 \log_2 d$
bits per channel use.

\section{Conclusions}
\label{sec:conc} 

In this work we have studied the behaviour of a fully correlated amplitude 
damping channel for two qubits. We assumed
that relaxation processes in the two qubits are strongly correlated, 
namely they only occur simultaneously
for the two qubits.  
We have considered three types of scenarios, 
the transmission of classical information, of quantum information and the
use of the channel in an entanglement-assisted fashion. We have derived 
the corresponding capacities (limiting to the single-shot capacity in 
the classical case), analytically studying the related maximization problems and 
individuating the optimal sources. In the case of classical capacity we 
also discussed the role of entanglement in achieve 
the maximum of Holvevo quantity. 

We find that the fully correlated amplitude damping channel 
is an interesting example of transmission
of classical or quantum information through a quantum channel 
for which a subspace is noiseless. Since the capacity $C_1$
is obtained without involving any coherent superposition 
of states from the noiseless and the noisy subspace, it
would be interesting to determine whether such result is 
specific for this model or more general.

A natural extension of our work would be to consider the case 
of amplitude damping channels with partial 
memory, i.e., $\mu<1$ in Eq.~(\ref{eq:model}). While the 
analytical solution of such model appears difficult, non trivial
bounds on the channel capacities could be computed.

\begin{acknowledgments}
A.D. and G.F. acknowledge support by PON02-00355-339123 - ENERGETIC, and by
EU through grant no. PITN-GA-2009-234970.
G.B. acknowledges the support by MIUR-PRIN project
``Collective quantum phenomena: From strongly correlated systems to 
quantum simulators''. 

\end{acknowledgments}

\appendix

\section{Optimality of the entangled ensembles for classical capacity}
\label{appx:ent}

Let us consider an ensemble ${\cal C}_s=\{{p}_k,\ket{{\psi}_k}\}$ 
of separable states
%which we \textit{design to achieve the bound} $C_1$:
\begin{eqnarray}
   \ket{\psi_k}&&=\,a_k \ket{00}\,+\,b_k\ket{01}\,+\,c_k\ket{10}\,+\,d_k\ket{11}\,=
   \label{eq:purestate-ensemble-inputstate}\\
               &&=\big(g_k\ket{0}+\sqrt{1-g_k^2}\ket{1}\big)\otimes\big(h_k\ket{0}+\sqrt{1-h_k^2}\ket{1}\big),\nonumber
\end{eqnarray}
where we can consider $g_k, h_k\in\mathbb{R}$
(since, as shown in (\ref{eq:eigenvalues-output-purestateE2}),
phases would not change the eigenvalues of the output state),
$g_k,\,h_k \in [0,1]$, and
such that the average density matrix is diagonal:
\begin{equation}
\rho = \left(
\begin{array}{cccc}
  \alpha & 0 & 0 & 0  \\
  0 & \beta & 0 & 0  \\
  0 & 0 & \gamma & 0  \\
  0 & 0 & 0 & \delta  \\
  \end{array} \right),
\label{eq:density-matrix-separablestates}
\end{equation}
with
\begin{eqnarray}
 &&\hspace{-0.0cm} \alpha=\sum_k p_k a_k^2, \quad 
                   \beta= \sum_k p_k b_k^2, \nonumber\\
 &&\hspace{-0.0cm} \gamma=\sum_k p_k c_k^2,  \quad 
                   \delta=\sum_k p_k d_k^2,
\end{eqnarray}
and
\begin{eqnarray}
 &&\hspace{-1.2cm} a_k^2 \,=\, g_k^2\,h_k^2,  \hspace{1.5cm}
                   b_k^2 \,=\, g_k^2(1-h_k^2), \nonumber\\
 &&\hspace{-1.2cm} c_k^2 \,=\, (1-g_k^2) h_k^2,\hspace{0.65cm} 
                   d_k^2 \,=\, (1-g_k^2)(1-h_k^2).
\end{eqnarray} 
We want to demonstrate that for any such ensemble, %which we design to reach $C_1$, 
we can find another ensemble ${\cal C}_e$, whose 
Holevo quantity is strictly greater than ${\cal C}_s$,
thanks to the presence of entangled states in ${\cal C}_e$.
We assume $\eta \,\in \, ]0,1[$, since we know that for the limiting
cases $\eta=0$ and $\eta=1$, an ensemble of separable state
succeeds in achieving $C_1$.

We start by considering that any such ensemble must have 
$\alpha, \beta, \gamma, \delta\neq0$. Indeed, since we are 
supposing that $\eta>0$, we know that $C_1>\log_2 3$ (see Fig. \ref{fig:bounds}),
therefore the entropy of (\ref{eq:density-matrix-separablestates})
has to be grater than $\log_2 3$, that is impossible
to achieve if even one of the parameters 
$\alpha, \beta, \gamma, \delta$ vanishes.
%$, while $\delta$ is equal to zero.
Next we subdivide ${\cal C}_s$ in two distinct subsets, 
${\cal C}_s\,=\,{\cal C}_{s_1}\cup{\cal C}_{s_2}$:
we collect all the states state with $(g_k=0,\,h_k=0)$ or
with $(g_k=1,\,h_k=1)$ in ${\cal C}_{s_2}$, all the others in 
${\cal C}_{s_1}$.

First we turn our attention to ${\cal C}_{s_1}$. We operate a substitution
similar to the one we applied at the 
beginning of Sec.~\ref{section:c1-capacity}.
We replace each state $\ket{\psi_{k_{s_1}}}$ and its occurrence probability $p_k$ 
in this ensemble by
\begin{eqnarray}
&&\hspace{-0.0cm} p_k,\quad\ket{\psi_{k_{s_1}}}\,\rightarrow\, \nonumber\\ \nonumber\\
&&\hspace{0.1cm}
   \left\{
   \begin{array}{ll}
    p_{\phi_k}=\frac{p_k (a_k^2+d_k^2)}{2}, &
               \ket{\phi_{k\pm}}=\frac{a_k}{\sqrt{a_k^2+d_k^2}}\ket{00}\pm
               \frac{d_k}{\sqrt{a_k^2+d_k^2}}\ket{11},\nonumber\\ %\label{eq:states-1}
               \quad & \quad \nonumber\\
               p_{\varphi 0}=p_k b_k^2, &
               \ket{\varphi_{k0}}=\ket{01},\nonumber\\
               \quad & \quad \nonumber\\
               p_{\varphi 1}=p_k c_k^2, &
               \ket{\varphi_{k1}}=\ket{10}.
               \end{array} \right. \\
               \label{eq:states-2fromsep}
\end{eqnarray}
It is straightforward to see that new ensemble, which we call ${{\cal C}}_{e_1}$,
has the same density matrix of ${\cal C}_{s_1}$, so it 
does not change the system output entropy.
With regard to 
the average output entropy we note that only states $\ket{\phi_{k\pm}}$ in 
${{\cal C}}_{e_1}$ contribute. The Holevo quantity for ensemble
${{\cal C}}_{e_1}$ is greater than for ${\cal C}_{s_1}$, 
since inequality (\ref{eq:lowerbound-outputentropy-c}), in which
we have to replace $2b_k^2 \to b_k^2 + c_k^2$, holds and is strict. 
As we have numerically verified, \textit{this is true except that for} 
$g_k=1,h_k\neq1$ or $g_k\neq1,h_k=1$  
(by construction we have excluded cases 
in which $g_k=1,h_k=1$), that is for $d_k=0$.

Now we turn to ensemble ${\cal C}_{s_2}$. Its density matrix is given by
\begin{equation}
\rho_{s_2} = \left(
\begin{array}{cccc}
  \alpha' & 0 & 0 & 0  \\
  0 & 0 & 0 & 0  \\
  0 & 0 & 0 & 0  \\
  0 & 0 & 0 & \delta'  \\
  \end{array} \right),
\label{eq:density-matrix-separablestates-2}
\end{equation}
where
\begin{eqnarray}
 &&\hspace{-1.2cm} \alpha'=\sum_{k\in{s_2}} p_k a_k^2, \quad 
                   \delta'=\sum_{k\in{s_2}} p_k d_k^2.\,  
%\qquad\textrm{where}\quad  a_k^2,\,d_k^2\,  \in \{0,1\}
\end{eqnarray}
We replace the ensemble ${\cal C}_{s_2}=\{\overline{p}_{\phi k},
\ket{\overline{\phi}_{k\pm}}\}$, by the following one, which we call 
${\cal C}_{e_2}$:
\begin{eqnarray}
               \overline{p}_{\phi k}=\frac{\alpha'+\delta'}{2},\quad
               \ket{\overline{\phi}_{k\pm}}=\sqrt{\frac{\alpha'}
{\alpha'+\delta'}}\ket{00}
                                \pm \sqrt{\frac{\delta'}{\alpha'+\delta'}}\ket{11}.\nonumber\\
               \label{eq:states-3fromsep}
\end{eqnarray}
The density matrix of ${\cal C}_{e_2}$ is equal to
(\ref{eq:density-matrix-separablestates-2}) and therefore
 the system output entropy does not change. 
Let we turn to the average output entropy. 
For the ensemble ${\cal C}_{s_2}$ it turns out that
\begin{eqnarray}
\overline{S}_{out,{\cal C}_{s_2}} \,=\,
\sum_{k \in {s_2}} {p}_k S({\cal E}_m(\ketbra{{\psi}_k}{{\psi}_k}))\,=\,\delta' H_2(\eta),
\end{eqnarray}
whereas for the ensemble ${\cal C}_{e_2}$ we have
\begin{eqnarray}
\overline{S}_{out,{\cal C}_{e_2}}
&&= (\alpha'+\delta') S({\cal E}_m(\ketbra{\overline{\phi}_{k\pm}}{\overline{\phi}_{k\pm}}))\nonumber\\
&&\hspace{-1cm}=(\alpha'+\delta') H_2\Bigg(\frac{1}{2}\Bigg[1+\sqrt{1-4\eta(1-\eta) \Big(\frac{\delta'}{\alpha'+\delta'}\Big)^2}\Bigg]\Bigg).\nonumber\\
\end{eqnarray}
Therefore, in order to show that replacing ${\cal C}_{s_2}$ 
with ${\cal C}_{e_2}$ we increase the Holevo quantity, 
we must prove that 
\begin{eqnarray}
&&\hspace{-0cm}\delta' H_2(\eta) \ge \nonumber\\
&&\hspace{+0.1cm}(\alpha'+\delta') H_2\Bigg(\frac{1}{2}\Bigg[1+\sqrt{1-4\eta(1-\eta) \Big(\frac{\delta'}{\alpha'+\delta'}\Big)^2}\Bigg]\Bigg).\nonumber\\
\label{eq:inequality-sep_ent-states}
\end{eqnarray}
We notice that the equality holds for $\alpha'\delta'=0$.
By dividing both members of 
(\ref{eq:inequality-sep_ent-states}) by $\delta'$ 
(assuming $\delta'>0$),
 inequality (\ref{eq:inequality-sep_ent-states})
is equivalent to
\begin{eqnarray}
H_2(\eta) \ge
x H_2\big(\frac{1}{2}\big[1+\sqrt{1-4\eta(1-\eta)x^{-2}}\big]\big),\,\, \forall \,x\in\, [1,\infty[.\nonumber\\
\label{eq_inequality77}
\end{eqnarray}
%It turns out that for $x=1$ (which corresponds to 
%the case $\alpha=0$) equality holds,
By numerical results it turns out that this inequality is tight 
for any $x>1$, that is, for any $\alpha'>0$, that together
with the previous assumption $\delta'>0$, and the fact the 
$\alpha'$ and $\delta'$ are populations, can be summarized 
as $\alpha'\delta'\neq0$.
%see fig. \ref{fig:proofTightnessInequality77}.
%\begin{figure}[ht]
%  \begin{center}
%  \includegraphics[width=10cm]{Inequality77}
%  \end{center}
%  \caption{Numerical evidence of the inequality (\ref{eq_inequality77})
%           as a function of $\eta \in [0,1]$ and $x \in [1,10^2]$ (we used
%           a logarithmic scale for $x$).  
%           The red surface represents the difference between
%           the two quantities appearing in the (\ref{eq_inequality77}). 
%           It is worth noting that the inequality is tight, except for $\eta \in\{0,1\}$ and $x=1$.}
%  \label{fig:proofTightnessInequality77}
%\end{figure} 

We can now conclude our proof that the ensemble 
${\cal C}_{e}={\cal C}_{e_1}\cup{\cal C}_{e_2}$ 
has a Holevo quantity strictly larger than ${\cal C}_{s}$. 
We observe that the two Holevo quantities can be written as
\begin{eqnarray}
 && \chi_{{\cal C}_{s}}=S(\rho)\,-\, \overline{S}_{out,{\cal C}_{s_1}}\,-\,\overline{S}_{out,{\cal C}_{s_2}},\nonumber\\
 && \chi_{{\cal C}_{e}}=S(\rho)\,-\, \overline{S}_{out,{\cal C}_{e_1}}\,-\,\overline{S}_{out,{\cal C}_{e_2}},\nonumber
\end{eqnarray}
since ${S}_{out,{\cal C}_{s}}={S}_{out,{\cal C}_{e}}=S(\rho)$ by construction.

As we must have $\delta\neq 0$, at least
one state in ${\cal C}_{s}$ has $d_k\neq 0$; we call this state $\ket{\xi}$. 
Suppose first $\ket{\xi}$ belongs to 
the subsets ${\cal C}_{s_1}$: we have already proved that 
$\overline{S}_{out,{\cal C}_{e_1}}<\overline{S}_{out,{\cal C}_{s_1}}$ (inequality
(\ref{eq:lowerbound-outputentropy-c})) and therefore 
$\chi_{{\cal C}_{e}}>\chi_{{\cal C}_{s}}$ 
(since in any case $\overline{S}_{out,{\cal C}_{e_2}}\le\overline{S}_{out,{\cal C}_{s_2}}$).
We can see that in this case the ensemble ${{\cal C}_{e_1}}$
must contain at least a pair of entangled states:
those states $\ket{\phi_{k\pm}}$ (\ref{eq:states-2fromsep}) corresponding to $\ket{\xi}$.
%the state in ${{\cal C}_{s_1}}$ with $d_k\neq0$. 
In fact, 
%the concerned state 
$\ket{\xi}$ must have $a_k\neq0$.
Actually in the case $a_k=0$ the inequality 
(\ref{eq:lowerbound-outputentropy-c}) implies that the ensemble 
${{\cal C}_{s_1}}$
has a Holevo quantity smaller than the one of ensemble ${{\cal C}_{e_1}}$;
in this case, ${{\cal C}_{e_1}}$ in turn exhibits a Holevo quantity 
of the form (\ref{eq:chi-A}), and we know that it does not achieve 
$C_1$ (see Fig.~\ref{fig:bounds}), so we have to discard
this case.
%, because ${{\cal C}_{s}}$ is designed to achieve $C_1$. 
If instead state $\ket{\xi}$ belongs to subset
${\cal C}_{s_2}$, we have to consider two further possibilities.
1) $\alpha'\neq 0$: inequality
(\ref{eq:inequality-sep_ent-states}) is tight and therefore 
$\overline{S}_{out,{\cal C}_{e_2}}<\overline{S}_{out,{\cal C}_{s_2}}$, 
which implies that $\chi_{{\cal C}_{e}}>\chi_{{\cal C}_{s}}$ 
(since in any case $\overline{S}_{out,{\cal C}_{e_1}}\le\overline{S}_{out,{\cal C}_{s_1}}$).
We stress that in this case the states in ${\cal C}_{e_2}$ are entangled.
2) $\alpha'= 0$: it is simple to verify that $C_s$ exhibits a Holevo 
quantity which is equal to the one of ensemble $\cal A$ 
(see eq. (\ref{eq:chi-A})), and
%, in which
%we have to replace $-2\beta\log_2\beta$ by $(-\beta\log_2\beta-\gamma\log_2\gamma)$);
$\chi_{{\cal C}_{s}}$ is strictly less than $C_1$, as one can see from Fig. \ref{fig:bounds},
so we can discard this case.

\section{Degradability of ${\cal E}_m$}
\label{appx:degradability}

We will consider a unitary representation of the channel ${\cal E}_m$
\begin{eqnarray}
&& \ket{00}^\textsf{S}\otimes\ket{00}^\textsf{E} \quad \longrightarrow \quad \ket{00}^S\otimes\ket{00}^\textsf{E} 
    \label{eq:unitary-representationEm}\\
&& \ket{01}^\textsf{S}\otimes\ket{00}^\textsf{E} \quad \longrightarrow \quad \ket{01}^S\otimes\ket{00}^\textsf{E} \nonumber\\
&& \ket{10}^\textsf{S}\otimes\ket{00}^\textsf{E} \quad \longrightarrow \quad \ket{10}^S\otimes\ket{00}^\textsf{E} \nonumber\\
&& \ket{11}^\textsf{S}\otimes\ket{00}^\textsf{E} \quad \longrightarrow \quad \sqrt{\eta}\ket{11}^\textsf{S}\otimes\ket{00}^\textsf{E}\,+\nonumber\\
&& \hspace{3.5cm}\,+\sqrt{1-\eta}\ket{00}^\textsf{S}\otimes\ket{11}^\textsf{E},\nonumber
\end{eqnarray}
where $\textsf{E}$ represents a fictitious environment.
When the system $\textsf{S}$ is prepared in the generic pure state (\ref{eq:generic-input-state}),
system $\textsf{SE}$ state undergoes the transformation
\begin{eqnarray}
&&\hspace{-0cm} \ket{\psi^{\textsf{SE}}}\,=\, \ket{\psi_k}^\textsf{S}\otimes\ket{00}^\textsf{E}\,\qquad \longrightarrow \nonumber\\
%&&\hspace{1.1cm}=\,a_k \ket{00}^\textsf{S}\otimes\ket{00}^\textsf{E}\,+\,b_k\ket{01}^\textsf{S}\otimes\ket{00}^\textsf{E}\,+\nonumber\\
%&&\hspace{1.5cm}  +\,c_k\ket{10}^\textsf{S}\otimes\ket{00}^\textsf{E}\,+\,d_k\ket{11}^\textsf{S}\otimes\ket{00}^\textsf{E} \qquad \longrightarrow\nonumber\\
\nonumber\\
&&\hspace{-0cm} \ket{\psi^{\textsf{SE}'}}\,=\      \,a_k \ket{00}^\textsf{S}\otimes\ket{00}^\textsf{E}\,+\label{eq:pure-state-transformation}\\
&&\hspace{1.5cm}+\,b_k\ket{01}^\textsf{S}\otimes\ket{00}^\textsf{E}\,+\,c_k\ket{10}^\textsf{S}\otimes\ket{00}^\textsf{E}\,+\nonumber\\
&&\hspace{1.5cm} +\,d_k\big(\sqrt{\eta}\ket{11}^\textsf{S}\otimes\ket{00}^\textsf{E}+\sqrt{1-\eta}\ket{00}^\textsf{S}\otimes\ket{11}^\textsf{E}\big).\nonumber
\end{eqnarray}
From (\ref{eq:pure-state-transformation}) we can calculate the reduced density
matrix for the systems $\textsf{S}$ and 
$\textsf{E}$;
%after the interaction (\ref{eq:unitary-representationEm}); 
$\rho'=\textrm{Tr}_{\textsf{E}}{\ketbra{\psi^{\textsf{SE}'}}{\psi^{\textsf{SE}'}}}$ is just
the output state (\ref{eq:output-purestateE2}), whereas the reduced density
matrix for the environment $\textsf{E}$ is 
\begin{eqnarray}
&&\hspace{-1cm}\rho^{\textsf{E}'}  =\textrm{Tr}_{\textsf{S}}{\ketbra{\psi^{\textsf{SE}'}}{\psi^{\textsf{SE}'}}}\nonumber\\
&&=\left(
\begin{array}{cccc}
  1-|d_k|^2(1-\eta)        & 0       & 0       & \sqrt{1-\eta}\,a_k d_k^*  \\
          0                & 0       & 0       & 0                       \\
          0                & 0       & 0       & 0                       \\
 \sqrt{1-\eta}\,a_k^* d_k    & 0       & 0       & (1-\eta) |d_k|^2 \\
  \end{array} \right).
\label{eq:ancilla-output-purestate} 
\end{eqnarray}
As we show in the following,
it is possible to deduce $\rho_{E'}$ starting from $\rho'$,
by applying to $\rho'$ a quantum operation and subsequently the channel 
${\cal E}_m$ in which he have to replace 
the parameter  $\eta$ by $(1-\eta)/\eta$.
This implies that the channel ${\cal E}_m$ is 
\textit{degradable}~\cite{degradable} for 
$\eta\,\in \,[\frac{1}{2},1]$.

In order to prove this we will consider a generic input state 
$\rho=\sum_k p_k\, \ketbra{\psi_k}{\psi_k}$, see
Eq.~(\ref{eq:generic-input-state}). the corresponding output state is give by
\begin{equation}
\rho'  =\left(
\begin{array}{cccc}
  \alpha+(1-\eta)\delta  & \kappa            & \lambda               & \sqrt{\eta}\, \varsigma  \\
    \kappa^*             & \beta             & \nu                   & \sqrt{\eta}\, o    \\
    \lambda^*            & \nu^*             & \gamma                & \sqrt{\eta}\, \pi  \\
  \sqrt{\eta}\,  \varsigma^*   & \sqrt{\eta}\,o^*  &  \sqrt{\eta}\, \pi^*  & \eta \delta        \\
  \end{array} \right),
\label{eq:output-genericstate} 
\end{equation}
and
\begin{equation}
\rho^{\textsf{E}'}  =\left(
\begin{array}{cccc}
  1-\delta(1-\eta)           & 0       & 0       & \sqrt{1-\eta}\,\varsigma  \\
          0                  & 0       & 0       & 0                   \\
          0                  & 0       & 0       & 0                   \\
 \sqrt{1-\eta}\,\varsigma^*        & 0       & 0       & (1-\eta) \delta     \\
  \end{array} \right),
\label{eq:ancilla-output-genericstate} 
\end{equation}
where:
\begin{eqnarray}
&& \alpha=\sum_k |a_k|^2, \,\,\, \beta=\sum_k |b_k|^2,\,\,\, \gamma=\sum_k |c_k|^2, \,\,\, \delta=\sum_k |d_k|^2, \nonumber\\
&& \kappa= \sum_k p_k a_k b_k^*,\quad \lambda=\sum_k p_k a_k c_k^*,\quad \varsigma=\sum_k p_k a_k d_k^*,  \nonumber\\
&& \nu=\sum_k p_k b_k c_k^*, \quad o=\sum_k p_k b_k d_k^*, \quad \pi=\sum_k p_k c_k d_k^*,
\end{eqnarray}
and moreover we have set 
\begin{eqnarray}
&& \alpha'=\alpha+(1-\eta)\delta, \quad \delta'=\eta \,\delta,\nonumber\\
&& \varsigma'=\sqrt{\eta}\,\varsigma, \quad o'=\sqrt{\eta}\,o, \quad \pi'=\sqrt{\eta}\,\pi.
\label{appx:positions}
\end{eqnarray}
%Now we can ask: "Is the channel ${\cal E}_m$ degradable"? That is: "Can we deduce 
%$\rho_{E'}$ starting from $\rho_{S'}$"?
To show that the channel ${\cal E}_m$ is degradable, we propose the following scheme.
We add three ancillary qubits to the system $\textsf{S}$ described 
by the state $\rho'$ (\ref{eq:output-genericstate});
we call the ancillas $\textsf{A}_1$ and $\textsf{A}_{23}$ (we collect
together the second and the third ancillary qubits). 
Initially the ancillas are all prepared in the state
$\ket{0}$. We first apply two Controlled NOT gates, where
the qubits $\textsf{S}$ act as control qubits and the 
qubit $\textsf{A}_1$ as the target qubit. We then perform a SWAP 
between $\textsf{S}$ and $\textsf{A}_{23}$, 
controlled by the state of the ancilla $A_1$.
This procedure is reported below
\begin{equation}
\begin{array}{ccc}
\textrm{Initial state}      & \longrightarrow  & \textrm{Controlled NOTs} 	\\	
 \ket{00^\textsf{S}}\otimes\ket{0^{\textsf{A}_1}} \otimes\ket{00^{\textsf{A}_{23}}} & \quad  &        \textrm{no changes}           \\ 
 \ket{01^\textsf{S}}\otimes\ket{0^{\textsf{A}_1}} \otimes\ket{00^{\textsf{A}_{23}}} &  \quad  & \ket{01^\textsf{S}}\otimes\ket{1^{\textsf{A}_1}} \otimes\ket{00^{\textsf{A}_{23}}} \\ 
 \ket{10^\textsf{S}}\otimes\ket{0^{\textsf{A}_1}} \otimes\ket{00^{\textsf{A}_{23}}} &  \quad  &\ket{10^\textsf{S}}\otimes\ket{1^{\textsf{A}_1}} \otimes\ket{00^{\textsf{A}_{23}}}  \\ 
 \ket{11^\textsf{S}}\otimes\ket{0^{\textsf{A}_1}} \otimes\ket{00^{\textsf{A}_{23}}} &   \quad  &      \textrm{no changes}            \\ 
  \end{array}\nonumber
\end{equation}
\normalsize
\small
\begin{equation}
\begin{array}{cc}
         \longrightarrow  &           \textrm{Controlled SWAP}\\
                          &            \textrm{no changes} \\ 
                          &            \ket{00^\textsf{S}}\otimes\ket{1^{\textsf{A}_1}} \otimes\ket{01^{\textsf{A}_{23}}} \\ 
                          &            \ket{00^\textsf{S}}\otimes\ket{1^{\textsf{A}_1}} \otimes\ket{10^{\textsf{A}_{23}}} \\ 
                          &            \textrm{no changes} \\ 
  \end{array}
\label{appx:OperationsWithAncilla}
\end{equation}
\normalsize

Exploiting the linearity of quantum operations we can transform each element
of $\rho^{S'}$ as
\begin{eqnarray}
&&\alpha'\,\ketbra{00}{00}\,\longrightarrow\nonumber\\
&&\hspace{2cm} \alpha'\,\ketbra{00}{00}\otimes\ketbra{0^{\textsf{A}_1}}{0^{\textsf{A}_1}}\otimes\ketbra{00^{\textsf{A}_{23}}}{00^{\textsf{A}_{23}}},\nonumber\\
&&\kappa \,\ketbra{00}{01}\,\longrightarrow\nonumber\\ 
&&\hspace{2cm} \kappa \,\ketbra{00}{00}\otimes\ketbra{0^{\textsf{A}_1}}{1^{\textsf{A}_1}}\otimes\ketbra{00^{\textsf{A}_{23}}}{01^{\textsf{A}_{23}}},\nonumber\\
%\end{eqnarray}
%\begin{eqnarray}
&&\lambda\,\ketbra{00}{10}\,\longrightarrow\nonumber\\
&&\hspace{2cm}\lambda\,\ketbra{00}{00}\otimes\ketbra{0^{\textsf{A}_1}}{1^{\textsf{A}_1}}\otimes\ketbra{00^{\textsf{A}_{23}}}{10^{\textsf{A}_{23}}},\nonumber\\
&&\varsigma'\,    \ketbra{00}{11}\, \longrightarrow\nonumber\\
&&\hspace{2cm} \varsigma'\, \ketbra{00}{11}\otimes\ketbra{0^{\textsf{A}_1}}{0^{\textsf{A}_1}}\otimes\ketbra{00^{\textsf{A}_{23}}}{00^{\textsf{A}_{23}}},\nonumber\\
&&\beta\,  \ketbra{01}{01}\,\longrightarrow\nonumber\\
&&\hspace{2cm} \beta\,\ketbra{00}{00}\otimes\ketbra{1^{\textsf{A}_1}}{1^{\textsf{A}_1}}\otimes\ketbra{01^{\textsf{A}_{23}}}{01^{\textsf{A}_{23}}},\nonumber\\
&&\nu\,    \ketbra{01}{10} \, \longrightarrow\nonumber\\
&& \hspace{2cm} \nu\,\ketbra{00}{00}\otimes\ketbra{1^{\textsf{A}_1}}{1^{\textsf{A}_1}}\otimes\ketbra{01^{\textsf{A}_{23}}}{10^{\textsf{A}_{23}}},
\nonumber\\
&& o'\,      \ketbra{01}{11} \, \longrightarrow\nonumber\\
&&\hspace{2cm} o'\,  \ketbra{00}{11}\otimes\ketbra{1^{\textsf{A}_1}}{0^{\textsf{A}_1}}\otimes\ketbra{01^{\textsf{A}_{23}}}{00^{\textsf{A}_{23}}},
\nonumber\\
&&\gamma\, \ketbra{10}{10}\,\longrightarrow\nonumber\\
&&\hspace{2cm}\gamma\,\ketbra{00}{00}\otimes\ketbra{1^{\textsf{A}_1}}{1^{\textsf{A}_1}}\otimes\ketbra{10^{\textsf{A}_{23}}}{10^{\textsf{A}_{23}}},\nonumber\\
&&\pi'\,    \ketbra{10}{11} \, \longrightarrow\nonumber\\
&& \hspace{2cm}\pi'\,\ketbra{00}{11}\otimes\ketbra{1^{\textsf{A}_1}}{0^{\textsf{A}_1}}\otimes\ketbra{10^{\textsf{A}_{23}}}{00^{\textsf{A}_{23}}},\nonumber\\
&&\delta'\, \ketbra{11}{11}\,\longrightarrow\nonumber\\
&&\hspace{2cm} \delta'\,\ketbra{11}{11}\otimes\ketbra{0^{\textsf{A}_1}}{0^{\textsf{A}_1}}\otimes\ketbra{00^{\textsf{A}_{23}}}{00^{\textsf{A}_{23}}}.\nonumber
\end{eqnarray}
\normalsize
After the quantum operations (\ref{appx:OperationsWithAncilla}), tracing with 
respect to the ancillas we obtain
\begin{eqnarray}
\rho''\, &=&\,  \alpha'\,\ketbra{00^\textsf{S}}{00^\textsf{S}}\,+\,\varsigma'\, \ketbra{00^\textsf{S}}{11^\textsf{S}}\,+\,{\varsigma'}^*\, \ketbra{11^\textsf{S}}{00^\textsf{S}}\,+\,\nonumber\\
 &&\hspace{0.3cm}     + \beta\,\ketbra{00^\textsf{S}}{00^\textsf{S}}\,+\,\gamma\,\ketbra{00^\textsf{S}}{00^\textsf{S}}\,+\delta'\,\ketbra{11^\textsf{S}}{11^\textsf{S}}
=\nonumber\\
&&\hspace{1.5cm}=\left(\begin{array}{cccc}
  \alpha'+\beta +\gamma & 0 & 0 &\varsigma' \nonumber\\
   0 & 0 & 0 & 0 \\
   0 & 0 & 0 & 0 \\
  {\varsigma'}^* & 0 & 0 & \delta' \\
  \end{array}\right)\,=\nonumber\\
&&\hspace{1.5cm}=\left(\begin{array}{cccc}
  1-\eta\,\delta & 0 & 0 &\sqrt{\eta}\,\varsigma \\
   0 & 0 & 0 & 0 \\
   0 & 0 & 0 & 0 \\
  \sqrt{\eta}{\varsigma}^* & 0 & 0 & \eta\,\delta \\
  \end{array}\right),
\label{eq:state-q}
\end{eqnarray}
where we have used (\ref{appx:positions}) together with $\alpha+\beta+\gamma+\delta=1$.
%since 
%\begin{eqnarray}
%&&\alpha'+\beta +\gamma=\alpha+\beta +\gamma+(1-\eta)\delta=\nonumber\\
%&&\hspace{1cm}        1-\delta+(1-\eta)\delta=1-\eta\delta\nonumber
%\end{eqnarray}
It is simple to see that one can obtain the state (\ref{eq:ancilla-output-purestate})
by applying  the channel ${\cal E}_m$ to the state (\ref{eq:state-q}), but
replacing $\eta$ with $(1-\eta)/\eta$. Indeed we have
\begin{eqnarray}
  \eta \delta &\quad\longrightarrow\quad& \eta\delta\cdot\frac{1-\eta}{\eta}=(1-\eta)\delta, \nonumber\\
  \sqrt{\eta} \varsigma &\quad\longrightarrow\quad& \sqrt{\eta} \varsigma \cdot\sqrt{\frac{1-\eta}{\eta}}=\sqrt{1-\eta}\,\mu, \nonumber\\
  1-\eta\delta &\quad\longrightarrow\quad& 1-\eta\delta+\Big(1-\frac{1-\eta}{\eta}\Big)\cdot\eta \delta=1-(1-\eta)\delta. \nonumber
\end{eqnarray}
It must of course happen that $0\le\frac{1-\eta}{\eta}\le 1$,
which means $\frac{1}{2}\le\eta \le1$.
We can therefore conclude that, when the transmissivity $\eta$ 
is in the interval $[\frac{1}{2},1]$, 
the channel ${\cal E}_m$ is degradable.


\begin{thebibliography}{99}

\bibitem{cover-thomas}
T. M. Cover and J. A. Thomas, 
\textit{Elements of Information Theory}
(Wiley, New York, 2006).

\bibitem{nielsen-chuang}
M. A. Nielsen and I. L. Chuang,
\textit{Quantum computation and quantum information}
(Cambridge University Press, Cambridge, 2000).

\bibitem{benenti-casati-strini}
G. Benenti, G. Casati, and G. Strini,
\textit{Principles of quantum computation and information}, vol. II
(World Scientific, Singapore, 2007).

\bibitem{hausladen}
P. Hausladen, R. Jozsa, B. Schumacher, M. Westmoreland, and W. K. Wootters, 
Phys. Rev. A \textbf{54}, 1869 (1996).

\bibitem{schumacher-westmoreland} 
B. Schumacher and M. D. Westmoreland,
%\textit{"Sending classical information via noisy quantum channels"}, 
Phys. Rev. A \textbf{56}, 131 (1997).

\bibitem{holevo98}
A. S. Holevo, IEEE Trans. Inf. Theory \textbf{44}, 269 (1998).

\bibitem{lloyd}
S. Lloyd, Phys. Rev. A \textbf{55}, 1613 (1997).

\bibitem{barnum}
H. Barnum, M. A. Nielsen, and B. Schumacher, 
Phys. Rev. A \textbf{57}, 4153 (1998).

\bibitem{devetak}
I. Devetak, IEEE Trans. Inf. Theory \textbf{51}, 44 (2005).

\bibitem{adami-Cerf} C. Adami and N. J. Cerf,
%\textit{"von Neumann capacity of noisy quantum channels"}, 
Phys. Rev. A \textbf{56}, 3470 (1997).

\bibitem{bennett1999}
C. H. Bennett, P. W. Shor, J. A. Smolin, and A. V. Thapliyal,
Phys. Rev. Lett. \textbf{83}, 3081 (1999).

\bibitem{bennett-shor}
C. H. Bennett, P. W. Shor, J. A. Smolin, and A. V. Thapliyal,
%\textit{"Entanglement-Assisted Capacity of a Quantum Channel and the Reverse Shannon Theorem"}, 
IEEE Trans. Inf. Theory \textbf{48}, 2637 (2002).

\bibitem{footnote}
The entanglement-assisted quantum capacity, i.e., the maximum 
amount of quantum information that can be sent through the 
channel per channel use with the help of prior unlimited entanglement, 
is simply given by $Q_E=C_E/2$; moreover, $Q\le Q_E$~\cite{bennett1999}.

\bibitem{banaszek}
K. Banaszek, A. Dragan, W. Wasilewski, and C. Radzewicz,
Phys. Rev. Lett. \textbf{92}, 257901 (2004).

\bibitem{solid-state}
Y. Makhlin, G. Sch\"on, and A. Shnirman,
Rev. Mod. Phys. \textbf{73}, 357 (2001);
E. Paladino, L. Faoro, G. Falci, and R. Fazio,
Phys. Rev. Lett. \textbf{88}, 228304 (2002);
G. Falci, A. D'Arrigo, A. Mastellone, and E. Paladino,
Phys. Rev. Lett. \textbf{94}, 167002 (2005);
G. Ithier, E. Collin, P. Joyez, P.J. Meeson, D. Vion, D. Esteve,
F. Chiarello, A. Shnirman, Y. Makhlin, J. Schriefl, and G. Sch\"on,
Phys. Rev. B \textbf{72}, 134519 (2005).

\bibitem{memo_review}
For a recent review, see 
F. Caruso, V. Giovannetti, C. Lupo, and S. Mancini,
preprint arXiv:1207.5435 [quant-ph].

\bibitem{mp02}
C. Macchiavello and G. M. Palma,
Phys. Rev. A \textbf{65}, 050301(R) (2002).

\bibitem{MMM}
L. Memarzadeh, C. Macchiavello and S. Mancini, 
New J. Phys.  {\bf 13}, 103031 (2011).

\bibitem{mpv04}
C. Macchiavello, G. M. Palma, and S. Virmani,
Phys. Rev. A \textbf{69}, 010303(R) (2004).

\bibitem{daems}
D. Daems,
Phys. Rev. A \textbf{76}, 012310 (2007).

\bibitem{dc}
Z. Shadman, H. Kampermann, D. Bruss and C. Macchiavello, 
Phys. Rev. A  {\bf 84}, 042309 (2011); Z. Shadman, H. Kampermann, D. Bruss 
and C. Macchiavello, Phys. Rev. A  {\bf 85}, 052306 (2012).

\bibitem{hamada}
H. Hamada, J. Math. Phys. \textbf{43} 4382 (2002),

\bibitem{njp}
A. D'Arrigo, G. Benenti, and G. Falci,
New J. Phys. \textbf{9}, 310 (2007).

\bibitem{virmani}
M. B. Plenio and S. Virmani,
Phys. Rev. Lett. \textbf{99}, 120504 (2007);
New J. Phys. \textbf{10}, 043032 (2008).

\bibitem{gabriela}
G. Barreto Lemos and G. Benenti,
Phys. Rev. A \textbf{81}, 062331 (2010).

\bibitem{lidar}
N. Arshed, A. H. Toor, and D. A. Lidar,
Phys. Rev. A \textbf{81}, 062353 (2010). 

\bibitem{cerf}
N.J. Cerf, J. Clavareau, C. Macchiavello and J. Roland,
Phys. Rev. A {\bf 72}, 042330 (2005).

\bibitem{mancini}
O. V. Pilyavets, V. G. Zborovskii, and S.Mancini,
Phys. Rev. A \textbf{77}, 052324 (2008).

\bibitem{lupo}
C. Lupo, V. Giovannetti, and S. Mancini,
Phys. Rev. Lett. \textbf{104}, 030501 (2010).

\bibitem{spins}
A. Bayat, D. Burgarth, S. Mancini, and S. Bose,
Phys. Rev. A \textbf{77}, 050306(R) (2008).

\bibitem{collision}
V. Giovannetti and G. M. Palma,
Phys. Rev. Lett. \textbf{108}, 040401 (2012). 

\bibitem{caruso}
F. Caruso, S. F. Huelga, and M. B. Plenio,
Phys. Rev. Lett. \textbf{105}, 190501 (2010).

\bibitem{micromaser}
G. Benenti, A. D'Arrigo, and G. Falci,
Phys. Rev. Lett. \textbf{103}, 020502 (2009);
A. D'Arrigo, G. Benenti, and G. Falci,
Eur. Phys. J. D \textbf{66}, 147 (2012).

\bibitem{bm04}
G. Bowen and S. Mancini,
Phys. Rev. A \textbf{69}, 012306 (2004).

\bibitem{datta}
N. Datta and T. C. Dorlas,
J. Phys. A: Math. Theor. \textbf{40}, 8147 (2007).

\bibitem{lmm09}
C. Lupo, L. Memarzadeh, and S. Mancini,
Phys. Rev. A \textbf{80}, 042328 (2009).

\bibitem{hastings}
M. B. Hastings, Nature Physics \textbf{5}, 255 (2009) 

\bibitem{yeo}
Y. Yeo and A. Skeen,
Phys. Rev. A \textbf{67}, 064301 (2003).

\bibitem{Jahangir} 
R. Jahangir, N. Arshed, and A. H. Toor,
%\textit{Quantum capacity of an amplitude-damping channel with memory}, 
preprint arXiv:1207.5612.

\bibitem{holevo73}
A. S. Holevo, Probl. Inf. Transm. \textbf{9}, 177 (1973).

\bibitem{dorlas-morgan08}
T. C. Dorlas and C. Morgan, Int. J. Quantum Inf.
\textbf{6}, 745 (2008).

\bibitem{giovannetti} 
V. Giovannetti and R. Fazio,
%\textit{"Information descritption of spin-chain correlation"}, 
Phys. Rev. A \textbf{71}, 032314 (2005).

\bibitem{schumachernielsen}
B. W. Schumacher and M. A. Nielsen, 
Phys. Rev. A \textbf{54}, 2629 (1996).

\bibitem{schumacher}
B. W. Schumacher, Phys. Rev. A \textbf{54}, 2614 (1996).

\bibitem{degradable}
I. Devetak and P. W. Shor, Comm. Math. Phys. \textbf{256}, 287 (2005).

\bibitem{wolf2007}
M. M. Wolf and D. P\'erez-Garc\'{\i}a, Phys. Rev. A \textbf{75}, 
012303 (2007).

\end{thebibliography}
\end{document}